\documentclass[a4paper]{article}
\usepackage[utf8]{inputenc}
\usepackage{amsmath, amsfonts, amsthm, amssymb, bbm, bm}
\usepackage{mathrsfs}
\usepackage{centernot}
\usepackage{enumerate,enumitem}
\usepackage{tabularx}
\usepackage{multicol}
\usepackage{multirow}
\usepackage[colorlinks=true,linktoc=all, allcolors=black]{hyperref}
\usepackage{setspace}
\usepackage[margin=3cm]{geometry}
\usepackage[round]{natbib}
\usepackage{booktabs}
\usepackage{tikz}
\usetikzlibrary{shapes,arrows,calc,fit}
\usepackage{algorithm}
\usepackage{algpseudocode} 
\usepackage{authblk}
\usepackage[english]{babel} 
\usepackage{blindtext}
\usepackage{lscape}
\usepackage{adjustbox,makecell}
\usepackage{listings,inconsolata}
\usepackage{xcolor} 
\renewcommand{\Pr}{\mathbb{P}}

\DeclareMathOperator{\Exp}{Exp}
\DeclareMathOperator{\expit}{expit}

\newcommand{\X}{\mathfrak{X}}

\DeclareMathOperator{\Bernoulli}{Bernoulli}

\newcommand{\cmid}{\,|\,}

\newcommand\indep{\protect\mathpalette{\protect\independenT}{\perp}}
\def\independenT#1#2{\mathrel{\rlap{$#1#2$}\mkern2mu{#1#2}}}

\renewcommand{\X}{Z}

\renewcommand{\L}{L}

\newcommand{\A}{A}

\theoremstyle{plain}
\newtheorem{lem}{Lemma}[section]

\newtheorem{proposition}[lem]{Proposition}

\theoremstyle{definition}

\newtheorem{remark}[lem]{Remark}

\newcommand{\benum}{\begin{enumerate}}
\newcommand{\eenum}{\end{enumerate}}

\newcommand{\bitem}{\begin{itemize}}
\newcommand{\eitem}{\end{itemize}}

\newcommand{\barr}{\begin{array}}
\newcommand{\earr}{\end{array}}

\newcommand{\bmat}{\begin{pmatrix}}
\newcommand{\emat}{\end{pmatrix}}

\newcommand{\blist}{\renewcommand{\labelenumi}{\textbf{\arabic{enumi}}.} \begin{enumerate}}
\newcommand{\elist}{\end{enumerate} \renewcommand{\labelenumi}{\arabic{enumi}.}}

\newcommand{\bs}{\boldsymbol}

\def\bal#1\eal{\begin{align*}#1\end{align*}}

\newcolumntype{L}{>{$}l<{$}}
\newcolumntype{C}{>{$}c<{$}}

\usepackage{tikz}
\usetikzlibrary{shapes, arrows, arrows.meta, calc, positioning}

\tikzset{nv/.style={circle, color=red, fill=red, inner sep=0.5mm}}
\tikzset{rv/.style={circle, draw, thick, minimum size=7mm, inner sep=0.5mm}}
\tikzset{fv/.style={rectangle, draw, thick, minimum size=7mm, inner sep=0.5mm}}
\tikzset{lv/.style={circle, color=red, fill=gray!30, draw, thick, minimum size=7mm, inner sep=0.5mm}}
\tikzset{rve/.style={ellipse, draw, thick, minimum size=7mm, inner sep=0.5mm}}

\tikzset{rvs/.style={circle, draw, thick, minimum size=6mm, inner sep=0.5mm}}
\tikzset{rvs0/.style={circle, minimum size=6mm, inner sep=0.5mm}}
\tikzset{fvs/.style={rectangle, draw, thick, minimum size=6mm, inner sep=0.5mm}}
\tikzset{lvs/.style={circle, color=red, fill=gray!30, draw, thick, minimum size=6mm, inner sep=0.5mm}}
\tikzset{rves/.style={ellipse, draw, thick, minimum size=6mm, inner sep=0.5mm}}

\tikzset{deg/.style={->, very thick, color=blue}}
\tikzset{degl/.style={->, very thick, color=red}}
\tikzset{beg/.style={<->, very thick, color=red}}
\tikzset{cdeg/.style={{Circle[length=+2pt 2.5,width=+2pt 2.5, fill=none]}->, very thick, color=blue}}
\tikzset{cceg/.style={{Circle[length=+2pt 2.5,width=+2pt 2.5, fill=none]}-{Circle[length=+2pt 2.5,width=+2pt 2.5, fill=none]}, very thick}}
\tikzset{uceg/.style={{Circle[length=+2pt 2.5,width=+2pt 2.5, fill=none]}-, very thick}}
\tikzset{ueg/.style={very thick}}

\definecolor{oxblue}{RGB}{0, 33, 71}
\usepackage{comment}

\title{Exact Simulation of Longitudinal Data from\\ Marginal Structural Models}

\author[1]{Xi Lin\thanks{Corresponding author: \texttt{xi.lin@stats.ox.ac.uk}.}}
\author[1]{Daniel de Vassimon Manela}
\author[2]{Chase Mathis}
\author[3]{\\Jens Magelund Tarp}
\author[1]{Robin J.~Evans}
\affil[1]{Department of Statistics, University of Oxford, UK}
\affil[2]{Department of Statistics, Duke University, USA}
\affil[3]{Novo Nordisk, Denmark}
\date{\today}

\renewcommand{\Tilde}{\widetilde}

\begin{document}

\maketitle

\onehalfspacing




\begin{abstract}

Simulating longitudinal data from specified marginal structural models is a crucial but challenging task for evaluating causal inference methods and informing study design. While data generation typically proceeds in a fully conditional manner using structural equations according to a temporal ordering, it is difficult to ensure alignment between conditional distributions and the target marginal causal effects, which presents a fundamental challenge. To address this, we propose a flexible and efficient algorithm for simulating longitudinal data that adheres exactly to a specified marginal structural model. Our approach accommodates time-to-event outcomes and extends naturally to survival settings, which are prevalent in applied research. Compared to existing approaches, it offers several advantages: it enables exact simulation from a known causal model rather than relying on approximations; avoids restrictive assumptions about the data-generating process; and remains computationally efficient by requiring only the evaluation of analytical expressions, rather than Monte Carlo methods or numerical integration. Through simulation studies replicating realistic scenarios, we validate the method's accuracy and utility. Our method will facilitate researchers in effectively simulating data with target causal structures for their specific scenarios.
\end{abstract}


\section{Introduction}\label{sec:intro}
Longitudinal data are fundamental for studies analyzing the causal effect of a treatment or exposure, especially for research into chronic diseases such as diabetes or HIV. Unlike cross-sectional studies, longitudinal designs capture the dynamic nature of treatment regimens, which often vary over time due to clinical decisions, patient adherence, or external factors. Even in clinical trials with strict adherence protocols, intercurrent events---such as treatment discontinuation or supplementary medication use---are common. 
By accounting for these treatment dynamics, longitudinal data enable the estimation of meaningful causal quantities, such as the ``on-treatment” estimand, which quantifies the treatment effect assuming treatment is administered as intended.

A central challenge in analyzing longitudinal data is addressing time-varying confounding, where a time-dependent covariate predicts both the outcome and future treatments, while itself being influenced by past treatments. For example, in the case study by \cite{robins2000marginal} on the effect of zidovudine (AZT) treatment on HIV infection, the biomarker CD4 lymphocyte count is both affected by prior AZT treatment and is predictive of future therapy initiation and HIV progression. The presence of such time-varying confounding renders conventional adjustment methods, like regression and stratification, generally biased.

Marginal structural models, introduced by \citet{robins1997msm}, have emerged as a powerful tool for estimating causal effects under time-varying confounding. Model parameters can be consistently estimated using inverse probability of treatment weighting \citep{robins1999association,robins2000marginalvs}. Among alternative methods, including structural nested models estimated through g-estimation \citep{robins1989analysis, robins1992estimation,robins1991correcting},  marginal structural models have gained prominence due to their interpretability and ease of implementation. A systematic review by \cite{clare2019causal} highlights their widespread use in applied research.


As causal inference methods for longitudinal data are continually proposed to address complex real-world settings, simulation studies are essential to evaluate their effectiveness for practical use. These studies assess important properties such as bias, confidence interval coverage, relative efficiency, and robustness to violations of the assumptions used, guiding both methodological development and application. Sample size calculations also benefit from simulations. Prior to conducting a trial or an observational study, it is crucial to determine the number of participants or observed units required to achieve adequate statistical power.
Although standard formulas can sometimes be used for time-fixed treatments, more complex scenarios with time-varying treatments make deriving analytical formulas impractical or impossible. In such cases, simulation is invaluable, allowing researchers to estimate sample sizes based on realistic data-generating processes and causal structures.

A fundamental requirement for both method evaluation and sample size determination is the ability to simulate data that conform to a specified causal model. However, simulating longitudinal data from a pre-specified marginal structural model is not a straightforward task. While generating data sequentially from structural equations---creating covariates, treatment status, and then the outcome conditioned on past information---may seem intuitive, these models rely on conditional rather than marginal specifications. This makes it difficult to sequentially simulate from conditional distributions that align with a pre-specified marginal model; in particular, it is extremely challenging to sample from a null hypothesis---the so-called \textit{g-null paradox }\citep{robins1997estimation}.

The remainder of this paper is structured as follows: Section \ref{sec:lit_review} reviews existing methods for simulating longitudinal data from causal models. Section \ref{sec:prelim} introduces key preliminaries, including the frugal parameterization \citep{Evans2024} and copulas. In Section \ref{sec:method}, we introduce our proposed methodology. Section \ref{sec:survival} extends the approach to time-to-event data, enabling simulations from survival models such as Cox marginal structural models, and Section \ref{sec:soul} illustrates the method through a simulation study replicating a real-world clinical scenario involving concomitant treatment in a trial. We conclude with a discussion in Section \ref{sec:discuss}. Example code for the simulations in Sections \ref{sec:illustration} and \ref{sec:soul} is available at \href{https://github.com/XiLinStats/LongSiMSM}{\texttt{https://github.com/XiLinStats/LongSiMSM}}.

\section{Related Literature}\label{sec:lit_review}

As noted in Section \ref{sec:intro}, the challenge lies in ensuring that the simulated data adhere to a specified \emph{marginal} model, rather than simply following sequential conditional distributions in a structural equation model. Existing approaches address this issue through two strategies:
\begin{enumerate}[label=(\roman*)]
    \item Simulating from conditional models and deriving or approximating the corresponding marginal models post hoc.
    \item Bypassing conditional simulations and simulating directly from the target marginal model.
\end{enumerate}


Several studies have explored the first approach. \cite{xiao2010accuracy} used sequential conditional models to approximate Cox marginal structural models under the rare disease assumption, i.e.,~very low failure rates, which mitigates the impact of non-collapsibility but assumes the absence of unmeasured confounding between time-varying confounders and outcomes. \cite{young2014simulation} employed the g-formula to bridge conditional and marginal models in discrete-time Cox marginal structural models. While theoretically feasible for discrete time-varying covariates, deriving marginal models becomes intractable for continuous covariates due to the need for numerical integration. Simplifications, such as the Markov property or rare disease assumptions, are often necessary but limit generalizability. \cite{sofrygin2017simcausal} developed the \texttt{simcausal} R package for flexible simulation from non-parametric structural equation models. The workflow involves specifying structural equations and sequentially simulating data from conditional distributions. However, deriving analytic expressions for marginal causal quantities under this conditional parameterization is often intractable and prone to model misspecification. To address this, users typically specify low-dimensional \textit{working} marginal structural models \citep{neugebauer2007nonparametric} and estimate their parameters via Monte Carlo simulations of counterfactual data. This introduces two limitations: first, the true marginal model remains unknown; second, reverse-engineering simulations to match a target causal function is challenging. \cite{keogh2021simulating} described simulation from additive-hazard marginal structural models \citep{aalen1989linear}. They utilized the result that if the conditional hazard model follows an additive form, the equivalent marginal hazard model will also be additive. While the marginal model retains the correct form, true parameter values are generally intractable and require Monte Carlo approximations. A further limitation is the difficulty in ensuring non-negative hazards for continuous covariates.

Alternative methods focus on simulating directly from the target marginal model. \cite{havercroft2012simulating} simulated directly from a discrete Cox marginal structural model, introducing time-varying confounding indirectly via a shared latent variable. However, this approach assumes no direct effect of time-varying confounders on outcomes, which may not hold in practice, e.g.,~the effect of CD4 count on HIV survival \citep{robins2000marginal}. \cite{Evans2024} proposed a \textit{frugal parameterization} that prioritizes the causal margin and parameterizes the surrounding variables accordingly, enabling direct simulation from the exact marginal model while explicitly modeling time-varying covariates. This addresses key limitations of earlier methods. Building on this, \cite{seaman2023simulating} applied the frugal parameterization to simulate survival outcomes, using a one-dimensional ``risk score'' to summarize multi-dimensional covariates. While simplifying dependence structures, this method relies on computationally expensive Monte Carlo approximations to estimate the cumulative distribution functions of the risk score. In settings with continuous treatments or baseline variables, these approximations must be performed separately for each individual at each time point, substantially increasing the computational burden.

The aims for a new simulation approach therefore include the ability to: simulate exactly from a specified marginal structural model; include multiple time-varying covariates, treatments, outcomes, and censoring; have good control over their distributions; avoid restrictive assumptions about the data-generating process.  In addition, we would like it to scale well in the number of time-points and variables, and applicable in real-world settings.

\section{Preliminaries}\label{sec:prelim}
\subsection{The frugal parameterization} \label{sec:prelim_frugal}

As discussed in the previous section, the fundamental challenge in simulating data from longitudinal marginal structural models arises from the fact that, while the joint data distribution can be easily factorized into conditionals for simulation, the causal quantity of interest is a marginal distribution.  Earlier methods, as detailed in Section \ref{sec:lit_review}, tackled this by deriving the marginal distribution implied by the conditional factorization. However, \cite{Evans2024} advanced this area with the introduction of the frugal parameterization, which explicitly parameterizes the causal marginal and constructs the rest of the joint distribution around it. This development has made the simulation method presented in this paper possible.

The frugal parameterization decomposes the joint distribution $p(\bs l,a,y)$ into three separate pieces:
\begin{enumerate}[label=(\roman*)]
    \item the past: $p_{\bs L A}(\bs l, a)$---the joint distribution of covariates and treatment prior to observing outcomes; 
    \item the causal margin: $p^*_{Y\mid A}(y\cmid do(a))$---the causal margin of interest, set according to the marginal structural models relevant to the causal question;
    \item a dependency measure: $\phi^*_{Y\bs L\mid A}$---captures the dependency between the outcome $Y$ and the covariates $\bs L$ under intervention.
\end{enumerate}
A common choice of this dependency measure is a copula, which we elaborate on in the following subsection.

\subsection{Copulas }
We can think of continuous multivariate distributions as the joining together of the univariate marginals (or margins) and the dependence structure. This structure can be represented by a copula, which is a multivariate distribution function $C:[0,1]^d \rightarrow [0,1]$ with margins uniformly distributed on $(0,1)$. According to Sklar's theorem \citep[]{sklar1959fonctions,sklar1973random}, let $F$ be a multivariate distribution with margins $F_1,\ldots, F_d$. Then there exists a copula $C$ such that for all $(x_1,\ldots x_d)$ in $\mathbb R^d$,
\begin{align*}
    F(x_1,\ldots,x_d) = C \left \{F_1(x_1), \ldots, F_d(x_d) \right\} = C(u_1,\ldots, u_d),
\end{align*}
where 
$u_i = F_i(x_i)$, $i = 1,\ldots, d$. If $F_1(x_1), \ldots, F_d(x_d) $ are all continuous, then $C$ is unique and takes the form
\begin{align*}
C(u_1,\ldots,u_d) = F \{F_1^{-1}(u_1), \ldots, F_d^{-1}(u_d) \}, \qquad (u_1,\ldots,u_d) \in [0,1]^d.
\end{align*}
Otherwise, $C$ is only uniquely determined on $\operatorname{Ran}F_1 \times \cdots \times \operatorname{Ran}F_d$, where $\operatorname{Ran}F_i$ denotes the range of $F_i$.
The choice of copula can be considered independently of the univariate margins. 

\subsubsection{Pair-copula construction}

While there exist a wide variety of parametric bivariate copulas, the options for higher-dimensional copulas are limited, with Gaussian and Student t-copulas being common choices. Extending symmetric bivariate Archimedean copulas to multivariate settings is also possible but imposes strong restrictions, such as requiring the same level of dependence among all possible pairs. Building on the work of \cite{joe1996families}, \cite{bedford2001probability,bedford2002vines} introduced pair-copula constructions, a flexible and intuitive approach to constructing multivariate copulas by modeling multivariate data as a series of bivariate copulas. This hierarchical structure incorporates additional variables at each level, using pair-copulas as the fundamental building blocks. 

Following the general formula given in \cite{joe1997multivariate}, and assuming that the joint distribution of $(Y ,X_1,\ldots, X_d)$ is absolutely continuous with continuous margins, the conditional distribution of $Y$ given $d$ other variables can be decomposed into $d$ bivariate copula density functions, each of which can be specified flexibly:
 %
\begin{align}
    f(y\cmid \bs x) = c_{YX_j|\bs X_{-j}} \left \{F(y\cmid \bs x_{-j}),F(x_j \cmid \bs x_{-j}) \right \} \cdot f(y\cmid \bs x_{-j}), \label{eqn:pcc_basic}
\end{align}
where $\bs x$ is a $d$-dimensional vector; $x_j$ is an arbitrarily chosen component of $\bs x$, and $x_{-j}  := (x_1, \ldots, x_{j-1}, x_{j+1}, \ldots, x_d) $ denotes $\bs x$ 
with its $j$th entry removed. Applying (\ref{eqn:pcc_basic}) recursively to the conditional densities, we have:
\begin{align}
    f(y\cmid \bs x) &= f(y) \cdot \prod_{i=1}^d  c_{YX_{d+1-i} |\bs X_{1:{d-i}}}\{F(y\cmid \bs x_{1:{d-i}}),F(x_{d+1-i} \cmid \bs x_{1:{d-i}}) \}, \label{eqn:pcc_basic2}
\end{align}

where each copula in the product in (\ref{eqn:pcc_basic2}) is bivariate; this defines a \emph{pair-copula}. In principle, any ordering of the conditioned variables $x_1,x_2,\ldots,x_d$ can be selected, although some orderings may be easier to use than others. In our proposed method, we choose an  ordering that is convenient and efficient for simulation. 

In conclusion, under appropriate regularity conditions, a multivariate density can be expressed as a product of pair-copulas, acting on different conditional distributions.

\subsubsection{Simulating from pair-copula constructions} \label{sec:sim_pcc}

The pair-copula construction includes conditional distributions of the form $F(y\cmid \bs x)$.  Assuming that the joint distribution of $(Y ,X_1,\ldots, X_d)$ is absolutely continuous with continuous margins, and the bivariate copula $C_{YX_j \mid \bs X_{-j}}$ is differentiable,  \cite{joe1996families} demonstrated the following result:
\begin{align}
        F(y\cmid \bs x) = \frac{\partial C_{YX_j \mid \bs X_{-j}} \left \{F(y\cmid \bs x_{-j}),F(x_j \cmid \bs x_{-j}) \right \}}{\partial F(x_j \cmid \bs x_{-j}) }, \label{eqn:pcc_cdf}
\end{align}
%
where $j \in \{1,\ldots,d\}$. For notational convenience, we introduce the \textit{h-function} $h(u_1,u_2; \Theta) $ from \cite{aas2009pair}. Let $u_1$ and $u_2$ be random variables with a uniform distribution on $(0,1)$, for a sufficiently regular copula $C_{12}(u_1,u_2)$, the $h$-function is defined as 
$$ h(u_1,u_2;\Theta) := F(u_1\cmid u_2) = \frac{\partial }{\partial u_2}C_{12}(u_1,u_2 ;\Theta)$$
where the second argument in $h(\cdot; \Theta)$ is the conditional variable, and $\Theta$ represents the set of copula parameters. The $h$-function is effectively the conditional copula  $C(u_1\cmid u_2;\Theta)$. We further define $h^{-1}(u_1,u_2; \Theta)$, be the inverse of the $h$-function in the first variable $u_1$. Equivalently, $ h^{-1}(u_1,u_2;\Theta) := F^{-1}(u_1\cmid u_2)$.  

\begin{remark}
    Here we define a copula $C$ as \textit{sufficiently regular} if it has non-zero density everywhere on its domain. Under this condition, the associated $h$-functions exist and are invertible. For standard parametric families---including the Gaussian, Student-t, Clayton and Frank copulas---the analytical form of $h(\cdot)$ and its inverse $h^{-1}(\cdot)$ can be derived directly from their definitions. See Appendix C of \cite{aas2009pair}.
\end{remark}

The $h$-functions are foundational for simulating from $F(y\cmid \bs x)$. Similar to (\ref{eqn:pcc_basic2}), by imposing an ordering $1,\ldots,d$ on the components of $\bs x$, we can recursively apply $h$-functions to obtain:
\begin{align}
    F(y\cmid \bs x) &= h\{F(y\cmid \bs x_{1:{d-1}}),F(x_d \cmid \bs x_{1:{d-1}}), \Theta_{d}\} \notag \\
    &=  h[h \{F(y\cmid \bs x_{1:{d-2}}),F(x_{d-1} \cmid \bs x_{1:{d-2}}; \Theta_{d-1}\},F(x_d \cmid \bs x_{1:{d-1}}); \Theta_{d}] \notag \\
    & \ldots \notag \\
    &= h [\ldots h\{h(F(y),F(x_1); \Theta_{1}),F(x_2\cmid x_1); \Theta_{2}\},\ldots ,F(x_d \cmid \bs x_{1:{d-1}}); \Theta_{d}]. 
\end{align}
Intuitively, the bivariate copulas ``weave'' the dependence between the outcome and each of the covariates. Thus, given a value for $\bs X$ we can simulate $Y$ by first drawing a uniform $U \sim {\rm U}(0,1)$, and then evaluating:
\begin{align}
    F(Y) = h^{-1} [\ldots h^{-1}\{h^{-1}(U, \, F(X_1); \Theta_{1}),F(X_2\cmid X_1); \Theta_{2}\},\ldots, F(X_d\cmid X_1, \ldots, X_{d-1}); \Theta_d]. \label{eqn:pcc_sim}
\end{align}
This means that to simulate $Y$ given a $d$-dimensional vector $\bs X$, we recursively calculate the $d$ inverse $h$-functions, forming a pair-copula construction. This underpins our proposed simulation method.

\section{Simulation Algorithm}\label{sec:method}

\begin{figure}
  \begin{center}
  \begin{tikzpicture}
  [rv2/.style={rv, inner sep=0.1mm}, node distance=20mm, >=stealth]
  \pgfsetarrows{latex-latex};
\begin{scope}
    \node[rvs0] (Z0) {$\L_0$};
    \node[rvs0, below right of=Z0] (X0) {$A_0$};
    \draw[rounded corners] ($(Z0)+(-5mm,+4mm)$) rectangle ($(X0)+(5mm,-7mm)$);
    \node[rvs0, above right of=X0] (Z1) {$\L_1$};
    \node[rvs0, below right of=Z1] (X1) {$A_1$};
    \draw[rounded corners] ($(Z1)+(-5mm,+4mm)$) rectangle ($(X1)+(5mm,-7mm)$);
    \node[right of=Z1] (d1) {$\cdots$};
    \node[right of=X1] (d2) {$\cdots$};
    \node[rvs0, above of=d1, yshift=-5mm] (C) {$Z$};
    \node[rvs0, right of=d1] (ZT) {$\L_K$};
    \node[rvs0, right of=d2, xshift=-5mm] (XT) {$A_K$};
    \draw[rounded corners] ($(ZT)+(-5mm,+4mm)$) rectangle ($(XT)+(12mm,-17mm)$);
    \node[rvs0, below right of=XT, xshift=-5mm, yshift=0mm] (YT) {$Y$};
    \draw[deg, orange] (C) to[bend right=15] (Z0);
    \draw[deg, orange] (C) to[bend right=25] (X0);
    \draw[deg, dashed, blue!50] (Z0) -- (Z1);
    \draw[deg] (Z0) -- (X0);
    \draw[deg, dashed, blue!50] (X0) -- (Z1);
\draw[deg, dashed, blue!50] (Z0) -- (X1);

    \draw[deg] (Z1) -- (X1);
    \draw[deg, orange] (C) -- (X1);
    \draw[deg, dashed, blue!50] (X0) -- (X1);
    \draw[deg, orange] (C) to[bend right=20] (XT);
    \draw[deg, orange] (C) -- (Z1);
    \draw[deg, orange] (C) to[bend left=0] (ZT);
    \draw[deg, dashed, blue!50] (Z1) -- (d1);
    \draw[deg, dashed, blue!50] (X1) -- (d2);
    \draw[deg, dashed, blue!50] (d1) -- (ZT);
    \draw[deg, dashed, blue!50] (d2) -- (XT);
   \draw[deg] (ZT) -- (XT);   
    \draw[deg, orange] (C) to[bend left=35] (YT);
    \draw[deg, dashed, blue!50] (Z1) to[bend right=0] (YT);
    \draw[deg, dashed, blue!50] (Z0) to[bend right=10] (YT);
    \draw[deg, dashed, blue!50] (X1) -- (YT);
    \draw[deg, dashed, blue!50] (X0) to[bend right=10] (YT);
    \draw[deg] (XT) -- (YT);
    \draw[deg] (ZT) to[bend right=15] (YT);

  \end{scope}
    \end{tikzpicture}
 \caption{An example directed acyclic graph showing temporal relationships in a longitudinal discrete setting, with baseline covariates $Z$; treatments 
 $A_0, \ldots, A_K$; time-varying covariates $L_0, \ldots, L_K$ and an outcome $Y$ observed at the end of the K follow up intervals.  Given the topological ordering 
 $Z,L_0,A_0,L_1,A_1\ldots,L_K,A_K,Y$, any variable may
 depend upon any subset of the previous variables.}
  \label{fig:longi}
  \end{center}
\end{figure}
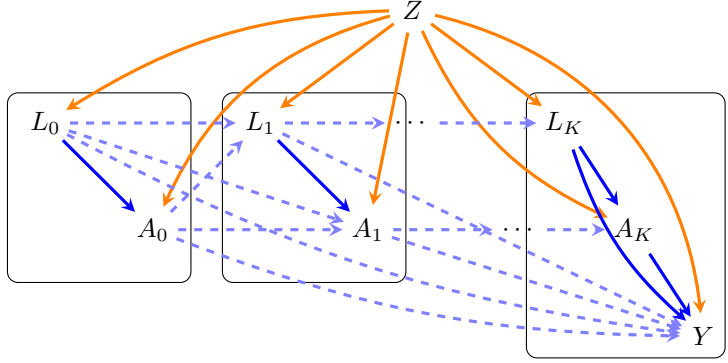
We first consider a longitudinal study with $n$ individuals observed at discrete time points, $\mathcal{K} = \{0,\ldots,K\}$. At each follow-up $k$, treatment status $A_k$, and a set of time-varying covariates $  L_k$ are recorded. Additionally, $Z$ represents a set of static baseline covariates. Latent variables, if present, are written as $\{H_k\}_{k \in \mathcal{K}}$. To indicate the history of a variable up to time $k$, we use bar notation. For instance, $\bar A_k = \{A_0,\ldots,A_k\}$ and  $\bar {L}_k = \{ L_0,\ldots,  L_k\}$, with the convention $\bar A_{-1} = \emptyset$ and $\bar L_{-1} = \emptyset$. The outcome $Y$ is observed at the final time point $K$.  Figure \ref{fig:longi} is a graphical representation of this setup, which is common in medical and epidemiological research. Examples include examining the effects of an HIV treatment on infection status assessed at the end of the study period \citep{robins2000marginal}, and assessing the impact of iron supplement use during pregnancy on the odds of anemia at delivery \citep{bodnar2004marginal}.
In Section \ref{sec:illustration}, we illustrate our method through a simulation study that replicates such settings.

Following the notation of \cite{Evans2024}, we use an asterisk to represent causal or interventional distributions where treatment is set to a specific value $A_k = a$ via intervention. In graphical terms, this corresponds to removing all arrows directed into $A_k$.

\subsection{The fully-conditional approach} \label{sec:fully_cond}

To outline our approach, we initially assume the time-varying covariate $L_k$ is one-dimensional and omit any latent process $\{H_k\}_{k \in \mathcal{K}}$; the extension to multi-dimensional time-varying covariates and latent variables is discussed in Appendix \ref{sec:app_latent}.

The standard method for simulation of longitudinal data $(Z, \bar{L}_K, \bar{A}_K, Y)$ is by time step in blocks using the Rosenblatt transformation \citep{rosenblatt1952remarks}. Specifically:
\begin{enumerate}
    \item \textbf{Baseline simulation}: Simulate static baseline variables, $z$ following a specified factorization, e.g.~ $F_{Z_1}, F_{Z_2\mid Z_1},\ldots$.
    \item \textbf{Time-step simulation}: At each time point $k=0,\dots,K$, simulate $ l_k = F_{L_k\mid Z, \bar{L}_{k-1},\bar{A}_{k-1}}^{-1}(w_{k,1},\cmid z, \bar{l}_{k-1}, \bar{a}_{k-1})$ and $a_k = F_{A_k\mid  Z, \bar{L}_{k},\bar{A}_{k-1}}^{-1}(w_{k,2},\cmid z, \bar{l}_{k}, \bar{a}_{k-1})$,
    where $w_{k,1},w_{k,2} \sim \operatorname{U}(0,1)$.
    \item \textbf{Outcome simulation}: Given the simulated covariates and treatments, generate the outcome $y = F_{Y\mid Z, \bar{L}_K,\bar{A}_K}^{-1}(w_y\cmid z,\bar{l}_{K}, \bar{a}_{K})$,
    with $w_{Y} \sim \operatorname{U}(0,1)$.
\end{enumerate}
Our proposed method diverges from this fully-conditional approach at Step 3. Instead of specifying the conditional distribution $F_{Y\mid Z \bar{L}_K\bar{A}_K}$---which inherently introduces misalignment between conditional and the desired marginal models---we introduce a margin-preserving parameterization for outcome simulation. 
\subsection{Frugal parameterization for outcome simulation}\label{sec:frugal_y}
Our method uses the frugal parameterization to simulate outcomes,
avoiding potentially intractable analytical calculations to obtain the marginal structural model.  
\cite{Evans2024} showed that the fully-conditional distribution of the outcome can be \textit{frugally} parameterized as:
\begin{align}
    p_{Y\mid \bar{L}_K,\bar{A}_K}(y\cmid \bar{l}_k,\bar{a}_K) = p^*_{Y\mid \bar{A}_K}(y\cmid do(\bar{a}_K)) \cdot c^*_{Y\bar{L}_K\mid \bar{A}_K}(u^*_Y, u_{L_0}, \dots, u_{L_K} \cmid do(\bar{a}_K)), \label{eqn:frugal_cond}
\end{align}
where $u^*_Y = F_{Y\mid \bar{A}_K}(y\cmid do(\bar{a}_K))$ and $u_{L_k} = F_{L_k\mid\bar{A}_K}(l_k\cmid do(\bar{a}_K))$ for $k=0,\ldots,K$ are the marginal quantiles under interventions that set $\bar{A}_K = \bar{a}_K$. 
\begin{remark}
    For brevity, we omit conditioning on baseline variables $ Z$ in all quantities in (\ref{eqn:frugal_cond}), treating them as if no baseline variables exist. This omission will continue hereafter, though all expressions can be extended to condition on $ Z$ by including it in their respective conditioning sets. 
\end{remark}
 This decomposition implies that we can simulate $Y$ through the multivariate copula and its interventional margin. Applying the pair-copula construction in (\ref{eqn:pcc_basic2}) to (\ref{eqn:frugal_cond}), we obtain:
\begin{align}
    p(y\cmid \bar{l}_K,\bar{a}_K) 
       &= p(y \cmid do(\bar{a}_K)) \cdot \prod_{k = 0}^{K} c^*\{F(y\cmid \bar{l}_{k-1}, do(\bar{a}_K)),F(l_k \cmid \bar{l}_{k-1},  do(\bar{a}_K)) ;\Theta_{k}\}. \label{eqn:outcome_non_survival}
\end{align}
Note that we omit some subscripts on densities for conciseness where doing so does not introduce ambiguity. This decomposes the conditional density into the product of $K+1$ pair-copulas and the target marginal causal density. The pair-copula linking $Y$ and $L_k$ is parameterized by $\Theta_k$, which may depend on intervened treatments $do(\bar{A}_k)$, past time-varying confounders $\bar{L}_{k-1}$ and baseline covariates $\bs Z$. Under intervention, graphically speaking, all arrows into nodes $\bar{A}_k$ are removed in the directed acyclic graph, rendering $L_k$ independent of subsequent interventions. Consequently, $F(l_k \cmid \bar{l}_{k-1}, do(\bar{a}_K)) = F(l_k \cmid \bar{l}_{k-1}, \bar{a}_{k-1})$ for $k = 1,\ldots, K$ and $F(l_0 \cmid do(\bar{a}_K)) = F(l_0)$. Following the simulation strategy from pair-copula constructions outlined in Section \ref{sec:sim_pcc}, the fully-conditional distribution of the outcome is alternatively parameterized as: 
\begin{align}
    F(y\cmid \bar{l}_K,\bar{a}_K)&= h \big [\dots h\big\{h(F_{Y\mid \bar{A}_K}^*(y\cmid  do(\bar{a}_K)), \,  F(l_0); \Theta_{0}), \, F(l_{1} \cmid l_0,  a_0)  ; \Theta_{1}\big\}\dots ; \Theta_{K} \big], \label{eqn:cond_cdf}
\end{align}
which is centered around the causal margin $F^*_{Y \mid \bar{A}_K}$.

We write $h_k(u_1,u_2) = h(u_1,u_2; \Theta_k)$, so as to avoid including the parameters $\Theta_k$. By the inversion method, it is straightforward that given ($\bar {l}_{K},\bar {a}_{K}$) and $U\sim \operatorname{U}(0,1)$,
    \begin{align*}
        Y &= F^{*-1}_{Y \mid \bar{A}_K}(h_0^{-1} \big[\dots h_{K-1}^{-1}\big\{h_K^{-1}(U, F(l_K \cmid \bar{l}_{K-1},\bar{a}_{K-1})),F(l_{K-1} \cmid \bar{l}_{K-2},\bar{a}_{K-2})\big\},\ldots, F(l_0) \big])
    \end{align*}
follows the distribution parameterized by  (\ref{eqn:outcome_non_survival}). 

    

This result means that following the approach outline in Section \ref{sec:fully_cond}, given the simulated baseline variables and time-varying variables, we can simulate the outcome in two steps: 
\begin{enumerate}
    \item Calculate the marginal quantile of the outcome, $u^*_Y$, as:
    \begin{align} 
        u^*_Y &= h_0^{-1} \big[\dots h_{K-1}^{-1}\big\{h_K^{-1}(w_Y,F(l_K \cmid \bar{l}_{K-1}, \bar{a}_{K-1})),  F(l_{K-1} \cmid \bar{l}_{K-2}, \bar{a}_{K-2})\big\}\dots, F(l_0 )\big]\label{eqn:sim_y}\\
        &= h_0^{-1} \big[\dots h_{K-1}^{-1}\big\{h_K^{-1}(w_Y, w_{K,1}),w_{K-1,1}\big\},\ldots, w_{0,1} \big], \nonumber 
    \end{align}
    where $w_Y \sim \operatorname{U}(0,1)$.
    \item Obtain $y$ through the inverse cumulative distribution function of the marginal distribution: $y = F^{*-1}_{Y \mid \bar{A}_K}(u^*_Y \cmid do(\bar{a}_K))$.
\end{enumerate}
An intuitive way to understand the calculation in Step 1 is that the inverse $h$-functions ``unravel'' the dependencies between $Y$ and each of $L_K, L_{K-1},\dots L_0$ sequentially, and remove the conditioning of $Y$ on $\bar{L}_K$ until the marginal quantile of $Y$ is obtained. The quantiles in the inverse $h$-functions, such as $F(l_k\cmid \bar{l}_{k-1},\bar{a}_{k-1})$, correspond exactly to the random standard uniform variables $w_{k,1}$ used to simulate $l_k$ at time step $k$. This enables direct substitution of the $l_k$ quantiles into (\ref{eqn:sim_y}) without additional computation. 

\subsection{Simulation Algorithm}
In summary, to simulate the data $( Z,\bar{L}_K, \bar{A}_K, Y)$, the user specifies:
\begin{itemize}
    \item the population distribution of the static baseline covariates $ Z$: $F_{ Z}$, either as a joint distribution, or through a conditional factorization;
    \item the development of the time-varying confounder $L_k$ as influenced by $\bs Z$ and $\bar{A}_{k-1}$: $F_{L_k\mid Z,\bar{L}_{k-1},\bar{A}_{k-1}}$;
    \item treatment assignment based on the time-varying confounder $\bar {L}_k$, past treatment  and baseline covariates $ Z$: $F_{A_k \mid \bs Z,\bar{L}_k,\bar{A}_{k-1}}$;
    \item the target causal marginal structural model: $F_{Y\mid \bar{A}_k,  Z}^{*}(\cdot\cmid do(\bar{a}_K), z)$;
    \item $K+1$ bivariate copulas describing dependencies between outcome $Y$ and $L_0, \ldots,L_K$: 
    $$C_{YL_k\mid Z, do(\bar{A}_k),\bar{L}_{k-1}}(\cdot; \Theta_k) \quad \text{for} \quad k=0,\ldots,K.$$
\end{itemize}

 Section \ref{sec:illustration} provides an example of specifying and interpreting these inputs through a simulation study that captures real-world complexities. 
 
 Algorithm \ref{alg:sim} below details the simulation procedure: 

\begin{algorithm}
\caption{Simulation algorithm for an individual $i$ over $K$ follow-ups}\label{alg:sim}
\begin{algorithmic}
    \State Sample $w_z \sim U(0,1)$
    \State  $z \leftarrow F_Z^{-1}(w_z)$
    \State Let $k = 0$
    \While{$k \leq K$}
        \State Sample $w_{k,1},w_{k,2}$ independent standard uniform variables
        \State $l_k \leftarrow F^{-1}_{L_k\mid Z,\bar{L}_{k-1},\bar{A}_{k-1}} (w_{k,1} \cmid z,\bar {l}_{k-1},\bar {a}_{k-1}) $
        \State $a_k \leftarrow F^{-1}_{A_k \mid Z,\bar{L}_k,\bar{A}_{k-1}}(w_{k,2}\cmid z,\bar {l}_k,\bar {a}_{k-1})$
        \State $k \leftarrow k+1$
        \EndWhile
        \State Sample $w_y \sim\operatorname{U}(0,1)$
        \State Let $\nu_{0} = w_{y}$
        \For{$j = 0$ to $K$}
            \State $\nu_{j+1} \leftarrow h_{K-j}^{-1}(\nu_{j},w_{(K-j),1})$
        \EndFor
        \State $u^*_y \leftarrow \nu_{K+1}$
        \State $y \leftarrow F_{Y\mid \bar{A}_k, Z}^{*-1}(u^*_{y}\cmid do(\bar{a}_K),z)$

    \State \Return $(z,\bar{l}_K,\bar{a}_K,y)$ 

\end{algorithmic}
\end{algorithm}

Extending the simulation algorithm to incorporate multiple time-varying confounders---whether assumed to be observed or unobserved --- is straightforward. The key modification lies in the outcome simulation: the marginal quantile of $Y$ is obtained by sequentially removing the conditioning on all the time-dependent confounders at each step, through inverse $h$-functions. We provide an illustrative example in Appendix \ref{sec:app_latent}, outlining the simulation procedure for incorporating an  additional latent process $\{H_k\}_{k\in\mathcal{K}}$.

\subsection{Illustrative example}\label{sec:illustration}
We demonstrate our simulation method by replicating the setting in \cite{bodnar2004marginal}, which examines the causal effect of iron supplement use during pregnancy on the odds of anemia at delivery. Participants are randomized into treatment and control groups and followed up over four visits leading up to delivery. Treatment is not time-fixed as participants are assumed to adhere to assigned treatment up to $k=2$, after which standard clinical protocols dictate further treatment based on hemoglobin levels.

The simulation considers baseline variables $ Z =\{$B = Pre-pregnancy obesity, C = Age$\}$ and time-varying covariate $L$ = $\{$Hemoglobin level (g/dL)$\}$. The time-dependent treatment, $A_k$, indicates whether an iron supplement is prescribed. The outcome $Y$ is measured at delivery and $Y=1$ indicates anemia at delivery.

The distribution of baseline variables is: $B\sim \Bernoulli(0.1), \, \, C \sim U(25,\,35), \, \, L_0 \sim N(11 -0.05\,B -0.02\, C, \,0.5)$ and $ A_0 \sim \Bernoulli(0.5)$.
For subsequent follow-up visits $k > 0$ we have $L_k \sim \operatorname{N}(L_{k-1} + 0.5\,A_{k-1},\,0.1)$, and $A_2 = A_1 = A_0$ with $A_k \sim \Bernoulli(\expit(10 -L_k + 0.1\,A_{k-1}))$ for $k = 3,4$.
We specify a logistic discrete-time hazard marginal structural model:
\begin{align*}
\Pr[Y = 1\cmid do(\Bar{A}_4),B,C] = \expit(-2 + 0.1\,B + 0.02 \,C - \beta \, \operatorname{cum}(\Bar{A}_4)),
\end{align*}
where $\operatorname{cum}(\Bar{A}_4) = \Sigma_{i=1}^{4}A_i$. Here, $\exp(\beta)$ represents the causal odds ratio for an incremental dose, and $\exp(4\beta)$ corresponds to the ``always treated” efficacy estimand.

We use a t-copula $T_{\rho,\operatorname{df}}$ (with $\rho = -0.5$ and $\operatorname{df}=5$) across all pair-copula construction components to model the dependency between $Y$ and each of $L_0, L_1,L_2,L_3$, and $L_4$. Compared to a Gaussian copula, the t-copula captures heavier-tailed dependency, capturing stronger associations in the tails reflecting stronger associations in extreme scenarios (e.g., severely low hemoglobin levels and anemia). Our algorithm offers full flexibility, allowing users to choose any copula family and specify different parameters for each bivariate copula in the pair-copula construction.

We simulated scenarios with varying causal coefficients $\beta = \{-0.5,-0.3,0 \}$, representing strong, moderate and no effect, respectively and samples sizes $n = \{500,1000,2000\}$. For each combination, we generated 400 datasets. The model $Y = \expit(\alpha + \gamma \,B + \theta \,C - \beta \, \operatorname{cum}(\Bar{A}_4))$ is fitted using stabilized inverse probability of treatment weights, and an unweighted approach. For each fit, we estimated 95\% confidence intervals for $\beta$ using both the robust sandwich estimator and non-parametric bootstrap.

Results are summarized in Table \ref{tab:demo1_bias}, \ref{tab:demo1_cov} and \ref{tab:demo1_power}. The inverse-probability-weighted estimator shows minimal bias, with magnitude well below Monte Carlo standard errors. In contrast, the unweighted estimator remains significantly biased across sample sizes, showing no convergence. Both the sandwich and bootstrap CIs of the inverse-probability-weighted estimator approach 95\% coverage, though with minor under-coverage. The unweighted estimator, however, exhibits poor coverage due to bias. Statistical power for testing the treatment effect (null hypothesis: $\beta = 0$) increases with effect size and sample size for the weighted estimator. In summary, these results confirm that data generated using our simulation algorithm accurately reflect the intended causal effects and time-varying confounding is introduced.

\begin{table}[h!]
    \begin{center}
         \begin{tabular}{cc cc}
        \toprule
        $\beta$ & $n$ & \textbf{IPTW} & \textbf{Unweighted} \\
        \midrule
        $-$0.5 & 500  & 0.027 (0.186) & 0.270 (0.115) \\
             & 1000 & 0.012 (0.140) & 0.269 (0.079) \\
             & 2000 & 0.009 (0.096) & 0.274 (0.055) \\
        \midrule
        $-$0.3 & 500  & 0.015 (0.179) & 0.236 (0.103) \\
             & 1000 & 0.004 (0.119) & 0.234 (0.073) \\
             & 2000 & 0.009 (0.088) & 0.235 (0.051) \\
        \midrule
         0   & 500  & 0.017 (0.138) & 0.189 (0.083) \\
             & 1000 & 0.015 (0.087) & 0.188 (0.059) \\
             & 2000 & 0.014 (0.072) & 0.193 (0.043) \\
        \bottomrule
    \end{tabular}
    \caption{Bias and Monte Carlo standard errors  (in parentheses, defined as $sd(\hat{\beta})$) for inverse-probability-of-treatment-weighted (`IPTW') and unweighted estimators (`unweighted') across different $\beta$ and $n$ scenarios.}
    \label{tab:demo1_bias}
    \end{center}
   
\end{table}

\begin{table}[h!]
    \centering
    \begin{tabular}{rr ccc c}
        \toprule
        \multicolumn{1}{c}{$\beta$} & \multicolumn{1}{c}{$n$} & \multicolumn{2}{c}{\textbf{IPTW}} & \textbf{Unweighted} \\
        \cmidrule(lr){3-4}
               &     & \textbf{Sandwich} & \textbf{Bootstrap} & \\
        \midrule
        $-0.5$ & 500  & 91.8\% & 93.0\% & 31.8\% \\
             & 1000 & 92.0\% & 91.8\% & 7.5\%  \\
             & 2000 & 92.3\% & 92.8\% & 0.0\%  \\
        \midrule
        $-0.3$ & 500  & 91.0\% & 91.5\% & 35.3\% \\
             & 1000 & 94.3\% & 92.5\% & 8.0\%  \\
             & 2000 & 94.3\% & 93.8\% & 0.3\%  \\
        \midrule
         0.0   & 500  & 92.8\% & 93.0\% & 41.8\% \\
             & 1000 & 95.5\% & 94.3\% & 12.0\% \\
             & 2000 & 92.8\% & 90.0\% & 0.5\%  \\
        \bottomrule
    \end{tabular}
    \caption{Coverage of 95\% confidence intervals for  IPTW-sandwich, IPTW-bootstrap, and unweighted estimators across scenarios with varying $\beta$ and $n$ values.}

    \label{tab:demo1_cov}
\end{table}

\begin{table}[h!]
    \centering
    \begin{tabular}{rr ccc c}
        \toprule
        \multicolumn{1}{c}{$\beta$}  & \multicolumn{1}{c}{$n$} & \multicolumn{2}{c}{\textbf{IPTW}} & \textbf{Unweighted} \\
        \cmidrule(lr){3-4}
               &     & \textbf{Sandwich} & \textbf{Bootstrap} & \\
        \midrule
        $-0.5$ & 500  & 83.3\% & 79.5\% & 55.3\% \\
             & 1000 & 97.5\% & 97.8\% & 82.5\% \\
             & 2000 & 100.0\% & 100.0\% & 98.3\% \\
        \midrule
        $-0.3$ & 500  & 45.8\% & 43.0\% & 12.0\% \\
             & 1000 & 80.0\% & 77.8\% & 15.3\% \\
             & 2000 & 95.8\% & 96.5\% & 26.8\% \\
        \midrule
         0.0   & 500  & 7.3\%  & 7.0\%  & 58.3\% \\
             & 1000 & 4.5\%  & 5.8\%  & 88.0\% \\
             & 2000 & 7.3\%  & 10.0\% & 99.5\% \\
        \bottomrule
    \end{tabular}
    \caption{Statistical power to reject the null hypothesis $\beta = 0$; that is, no causal effect.}
            \label{tab:demo1_power}
\end{table}


\section{Simulating Survival Outcomes}\label{sec:survival}
\begin{figure}
  \begin{center}
  \begin{tikzpicture}
  [rv2/.style={rv, inner sep=0.1mm}, node distance=20mm, >=stealth]
  \pgfsetarrows{latex-latex};
\begin{scope}
  \node[rvs0] (Z0) {$\L_0$};
  \node[rvs0, below right of=Z0, xshift=-6mm] (X0) {$\A_0$};
  \node[rvs0, above right of=X0] (Z1) {$\L_1$};
  \node[rvs0, below right of=Z1,xshift=-6mm] (X1) {$\A_1$};
  \node[rvs0, below right of=X0, xshift=-6mm, yshift=-5mm] (Y1) {$Y_1$};
  \node[right of=Z1] (d1) {$\cdots$};
    \node[right of=X1] (d2) {$\cdots$};
      \node[rvs0, below right of=X1, xshift=-6mm, yshift=-5mm] (Y2) {$Y_{2}$};
  \node[rvs0, above of=d1, yshift=-5mm] (C) {$Z$};
    \node[rvs0, right of=d1, xshift=-6mm] (ZT) {$\L_K$};
  \node[rvs0, right of=d2, xshift=-6mm] (XT) {$\A_K$};
  \node[rvs0, below right of=XT, xshift=-6mm, yshift=-5mm] (YT) {$Y_{K+1}$};
  \draw[deg, color=orange] (C) to[bend right=15] (Z0);
  \draw[deg, color=orange] (C) to[bend right=25] (X0);
  \draw[deg, color=orange] (C) to[bend right=10] (Y1);
  
  \draw[deg, dashed, blue!50] (Z0) -- (Z1);
  \draw[deg] (Z0) -- (X0);
  \draw[deg, dashed, blue!50] (X0) -- (Z1);
   \draw[deg] (X0) -- (Y1);
   \draw[deg] (Z0) to[bend right=15] (Y1);
     \draw[rounded corners] ($(Y1)+(-19mm,-4mm)$) rectangle ($(Y1)+(2mm,40mm)$);
  \draw[deg, dashed, blue!50] (Z0) -- (X1);
  \draw[deg] (Z1) -- (X1);
  \draw[deg, color=orange] (C) -- (X1);
  \draw[deg, color=orange] (C) -- (Y2);
  \draw[deg, dashed, blue!50] (X0) -- (X1);
     \draw[deg] (X1) -- (Y2);
   \draw[deg, dashed, blue!50] (Z0) to  (Y2);
      \draw[deg] (Z1) to[bend right=15] (Y2);
           \draw[rounded corners] ($(Y2)+(-19mm,-4mm)$) rectangle ($(Y2)+(2mm,40mm)$);
   \draw[deg, color=orange] (C) to[bend right=5] (XT);
   \draw[deg, color=orange] (C) -- (Z1);
   \draw[deg, color=orange] (C) to[bend left=0] (ZT);
   \draw[deg, dashed, blue!50] (Z1) -- (d1);
   \draw[deg, dashed, blue!50] (X1) -- (d2);
   \draw[deg, dashed, blue!50] (d1) -- (ZT);
   \draw[deg, dashed, blue!50] (d2) -- (XT);
   \draw[deg] (ZT) -- (XT);   
  \draw[deg, color=orange] (C) to[bend left=35] (YT);
     \draw[rounded corners] ($(YT)+(-20mm,-4mm)$) rectangle ($(YT)+(5mm,40mm)$);
     \draw[deg, dashed, blue!50] (X1) -- (ZT);
     \draw[deg, dashed, blue!50] (Z1) -- (XT);
  \draw[deg, dashed, blue!50] (Z1) to[bend right=0] (YT);
  \draw[deg, dashed, blue!50] (Z0) to[bend right=10] (YT);
  \draw[deg, dashed, blue!50] (X1) -- (YT);
  \draw[deg, dashed, blue!50] (X0) to[bend right=10] (YT);
    \draw[deg] (XT) -- (YT);
    \draw[deg] (ZT) to[bend right=15] (YT);

  \end{scope}
    \end{tikzpicture}
 \caption{Example directed acyclic graph showing temporal relationships in a longitudinal survival setting. The sequence of outcomes observed over time is denoted as $\{Y_k\}_{k \in \mathcal{K}^+}$, where $\mathcal{K}^ + = \{1,\ldots, K+1\}$.} 
  \label{fig:longi_surv}
  \end{center}
\end{figure}
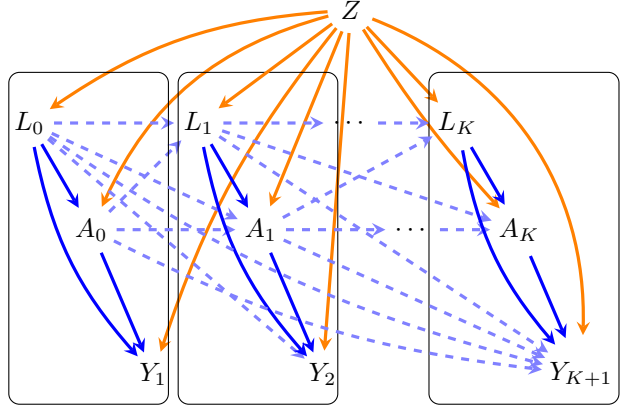

We have described and demonstrated our method to simulate longitudinal data where the outcome is measured at the final time point. Since time-to-event and survival outcomes are very common in medical and epidemiological contexts, we shall now demonstrate how our method can be extended to simulate longitudinal data from survival models.

In Figure \ref{fig:longi_surv}, each plate represents a single interval. For the $k$th interval, where $k = 0,\ldots, K$, the observations include $L_k, A_k$ at the start of the interval, and the survival status $Y_{k+1}$ at the end. The length of each interval is set to 1. In addition to the notation in Section \ref{sec:method}, let $T$ denote an individual's failure time. The outcome is represented be a sequence $\{Y_k\}_{k \in \mathcal{K}^+}$, where $\mathcal{K}^ + = \{1,\ldots, K+1\}$ and $Y_{k+1} = I(T < k+1)$ indicates failure. 

We describe the algorithm for the case where the time-varying confounder is univariate and no latent processes are present. The algorithm can be extended analogously to the case of ordinary longitudinal outcomes. The illustrative example at the end of this section demonstrates the simulation with two time-varying confounders.

Depending on the chosen marginal structural model, there are various ways to model the outcomes. Examples include:
\begin{itemize}
    \item Discrete-time marginal structural model, where the binary survival outcome $Y_k$ is modeled directly: 
    \begin{align}
        P(Y_{k+1} = 1 \cmid do(\bar{A}_k = \bar{a}_k), Z = z, Y_{k} = 0) = f(\bar{a}_k,z; \alpha), \label{eqn:disc_cox}
    \end{align}
    where $f$ can be any real-valued functions bounded between 0 and 1, for example, an expit function for a logistic marginal structural model.
    \item Continuous-time hazard MSM, where we introduce an auxiliary continuous variable, $\Tilde{Y}_{k+1} = T - k \cmid T >k $, representing the incremental failure time given survival at the start of the $k$th interval. We model incremental survival time $\Tilde{Y}_{k+1}$ and truncated it to obtain the observed binary survival outcome $Y_{k+1}$. Given survival at $k$, the time to failure is $T\cmid T>k = k +\Tilde{Y}_{k+1} $. Examples of marginal models for $\Tilde{Y}_{k+1}$ include:
        \begin{itemize}
            \item Cox proportional hazard: $$\lambda(\Tilde{y}_{k+1}\cmid do(\bar{a}_k), z,y_{k} = 0) = \lambda(\Tilde{y}_{k+1}\cmid do(\bar{0}_k), z,y_{k} = 0) \exp \{g(\bar{a}_k,z ; \beta)\},$$
            \item Additive hazard: $$\lambda(\Tilde{y}_{k+1}\cmid do(\bar{a}_k), z, y_{k} = 0) = \lambda(\Tilde{y}_{k+1}\cmid do(\bar{0}_k), z, y_{k} = 0) + g(\bar{a}_k,z ; \gamma).$$
        \end{itemize}
    
\end{itemize} 
We focus on describing the simulation algorithm for time-to-event outcomes, with a remark on adapting the approach for discrete survival outcomes, which follows a similar but simpler framework.

\subsection{Simulation algorithm} \label{sec:survivl_algo}

The general approach remains similar to the non-survival case, but with the key difference that at each time step $k$, simulation is now conditioned on survival up to the start of the current time interval. For brevity, we continue to omit conditioning on $Z$.

Unlike the derivation of outcome simulation for the non-survival case in Section \ref{sec:frugal_y}, all quantities involved in the copulas must now be conditioned on survival up to this point. Specifically, when simulating the incremental failure time, $\Tilde{y}_{k+1}$, given $Y_k = 0$, (\ref{eqn:outcome_non_survival}) becomes:
\begin{align}
    p(\Tilde{y}_{k+1}   \cmid \bar{l}_k,\bar{a}_k, y_k = 0) = & p^*(\Tilde{y}_{k+1}  \cmid do(\bar{a}_k), y_k = 0) \cdot \notag\\
      &  \prod_{j = 0}^{k} c^*_{\Tilde{Y}_{k+1}, L_{j} \mid \bar{L}_{j-1}, \bar{A}_k,  Y_k = 0}\big\{F(\Tilde{y}_{k+1}\cmid \bar{l}_{j-1}, do(\bar{a}_k),  y_k = 0)), \nonumber \\
      & \qquad\qquad\qquad\qquad\qquad\qquad\qquad F(l_j \cmid \bar{l}_{j-1},do(\bar{a}_{k}), y_k = 0) ;\Theta_{k,j}\}. \label{eqn:outcome_survival}
\end{align}

It is important to note the distinction between the subscripts $j$ and $k$ in the product. To reiterate, our goal is to simulate the outcome at the end of the $k$th interval, which corresponds to the $k$th panel in Figure \ref{fig:longi_surv}. Each copula in (\ref{eqn:outcome_survival}) associates $\Tilde{y}_{k+1}$ with all the prior values of the time-varying confounder $L_j$, for $j=0,\ldots,k$. 

In the non-survival case discussed in Section \ref{sec:frugal_y}, the quantiles $F(l_j \cmid \bar{l}_{j-1}, \bar{a}_{j-1})$, for $j=0,\ldots,k$, are directly available at each simulation at step $j$ and can be simply plugged into the iterative application of inverse $h$-functions without further computation. However, this convenience does not extend when the quantiles are conditioned on survival at later time points. For instance, suppose that at $k=0$, we simulate baseline health, $P(L_0) \sim \operatorname{N}(0,1)$. Conditioning on survival after 12 months, however, alters the distribution. The baseline health of the survivors will almost certainly not be representative of the initial population, meaning that the distribution is no longer standard normal. This means that the conditional quantiles of $L_j$ that we need are not directly available and require additional computation. Fortunately, employing copulas to define the dependence structure offers a solution to deriving these conditional distributions.
\begin{lem} In the setting described above, suppose $Y_k = 0 \Leftrightarrow\Tilde{Y}_k > 1$, then the conditional distribution $F_{L_j \cmid \bar{L}_{j-1},do(\bar{A}_{k}), Y_k = 0}$ is given by
        \begin{align}
    F(l_j \cmid \bar{l}_{j-1},do(\bar{a}_{k}), y_k = 0) &= \frac{v_{k-1,j} - C_{\Tilde{Y}_kL_j\mid \bar{L}_j,do(\bar{A}_k),Y_{k-1} = 0}(v_{k-1,j},q_{k-1,j};\Theta_{k-1,j})} {1-q_{k-1,j}},
\end{align}
where $v_{k-1,j} = F(l_j \cmid \bar{l}_{j-1},do(\bar{a}_{k}),y_{k-1} = 0)$ and $q_{k-1,j} =F(\Tilde{y}_k = 1 \cmid \bar{l}_{j-1},do(\bar{a}_{k}),  y_{k-1} = 0)$.

\end{lem}\label{lem}
The proof is given in Appendix \ref{sec:app_lemma}. Here we emphasize that $F_{L_j \cmid \bar{L}_{j-1},do(\bar{A}_{k}), Y_k = 0}$ is the ``renormalized'' distribution in the population who have survived up to time-point $k$. This is distinct from simply conditioning on an individual's previous incremental failure time, which would give $F_{L_j \cmid \bar{L}_{j-1},do(\bar{A}_{k}), \Tilde{Y}_k}$. The former corresponds to the expectation of the latter with respect to $\Tilde{Y}_k$ given $\Tilde{Y}_k >1$.



The proposed simulation algorithm combines an iterative and recursive structure and is formally presented and validated using an inductive approach.

\textbf{Proposed simulation method}: At time step $k$, we specify the cumulative distribution functions: 
\begin{align*}
    F_{L_k\mid \bar{L}_{k-1}, \bar{A}_{k-1}, Y_k = 0} && F_{A_k\mid \bar{L}_{k}, \bar{A}_{k-1}, Y_k = 0} && F_{\Tilde{Y}_{k+1}\mid do(\bar{A}_{k}), Y_k = 0}.
\end{align*}
Additionally, we define $k+1$ bivariate copulas $C_{\Tilde{Y}_{k+1}L_j\mid \bar{L}_{j-1},do(\bar{A}_k),Y_k = 0}$, parameterized by $\Theta_{k,j}$ for $j=0,\ldots,k$. These copulas captures the dependence between $\Tilde{Y}_{k+1}$ and each preceding time-varying covariate. Draw $w_{k,1}$, $w_{k,2}$, $w_{k,3} \sim \operatorname{U}(0,1)$ and compute:
\begin{align}
    l_k = F^{-1}_{L_k\mid \bar{L}_{k-1}, \bar{A}_{k-1}, Y_k = 0}(w_{k,1} \cmid  \bar{l}_{k-1}, \bar{a}_{k-1}) && a_k = F^{-1}_{A_k\mid \bar{L}_{k}, \bar{A}_{k-1}, Y_k =0}(w_{k,2} \cmid \bar{l}_{k}, \bar{a}_{k-1}),
\end{align}
to obtain a sample $(l_k,a_k)$. We initialize by setting $ F(\Tilde{y}_{k+1}\cmid \bar{l}_{k},do(\bar{a}_k), y_k = 0) = w_{k,3}$. Then, for $j = k-1,\ldots,0$, we iteratively apply the transformations:
 \begin{align}
         F(l_{j} \cmid \bar{l}_{j-1},do(\bar{a}_k),y_k =0 ) &= \frac{v_{k-1,j} - C_{\Tilde{Y}_kL_j\mid \bar{L}_j,do(\bar{A}_k),Y_{k-1} = 0}(v_{k-1,j},q_{k-1,j};\Theta_{k-1,j})} {1-q_{k-1,j}}, \label{eqn:ljyk}\\
        F(\Tilde{y}_{k+1}\cmid \bar{l}_{j},do(\bar{a}_k), y_k = 0) &= h^{-1}\big\{ F(\Tilde{y}_{k+1}\cmid \bar{l}_{j+1}, do(\bar{a}_k),y_k = 0), \nonumber\\
        &\qquad\qquad F(l_{j+1}\cmid \bar{l}_{j},do(\bar{a}_k), y_k = 0); \Theta_{k,j+1} \big\}. \label{eqn:ylk}
    \end{align}    
Expression (\ref{eqn:ljyk}) follows Lemma \ref{lem}, by induction, $v_{k-1,j}$ and $q_{k-1,j}$ are available from the previous time step $k-1$. 

When the recursion reaches $j = 0$, we obtain the quantile $F(\Tilde{y}_{k+1}\cmid l_0, do(\bar{a}_k), y_k = 0)$. Applying  the $h$-function one final time, we set:
\begin{align*}
    u^*_{k+1} = F(\Tilde{y}_{k+1}\cmid do(\bar{a}_k)) &= h^{-1}\big\{ F(\Tilde{y}_{k+1}\cmid l_{0}, do(\bar{a}_k),y_k = 0), \nonumber\\
        &\qquad\qquad F(l_{0}\cmid do(\bar{a}_k), y_k = 0); \Theta_{k,0} \big\}.
\end{align*}
A sample of $\Tilde{y}_{k+1}$ is then obtained via inversion:
\begin{align*}
    \Tilde{y}_{k+1} = F_{Y_{k+1}\mid do(\bar{A}_k),Y_k = 0}^{-1}(u^*_{k+1}), 
\end{align*}
with the corresponding binary outcome $y_{k+1} = I(\Tilde{y}_{k+1} <1) $. If $y_{k+1} = 1$, the simulation stops; otherwise, before proceeding to $k +1$, we store and compute the following quantities for the next step: 
\begin{itemize}
    \item Store $v_{k,j}$ = $ F(l_{j} \cmid \bar{l}_{j-1},do(\bar{a}_k),y_k =0 )$, for $j = 0,\ldots,k$ , which have been computed in the current step via (\ref{eqn:ljyk}).
    \item Compute $q_{k,j} =  F(\Tilde{y}_{k+1} = 1 \cmid \bar{l}_{j-1},do(\bar{a}_{k}),  y_k = 0)$: initialize 
    $q_{k,0} = F(\Tilde{y}_{k+1} = 1 \cmid do (\bar{a}_k), y_k = 0)$, then for $j = 1, \ldots k$, recursively update:
\begin{align}
    q_{k,j} &= F(\Tilde{y}_{k+1} = 1 \cmid \bar{l}_{j-1},do(\bar{a}_{k}),  y_k = 0) \nonumber \\
    & = h \big \{ F(\Tilde{y}_{k+1} = 1 \cmid \bar{l}_{j-2 },do(\bar{a}_{k}),  y_{k} = 0),F(l_{j-1} \cmid \bar{l}_{j-2 },do(\bar{a}_{k}),  y_{k} = 0);\Theta_{k,j-1}\big\}\nonumber\\
    &= h \big \{ q_{k,j-1},v_{k,j-1};\Theta_{k,j-1}\big\}. \label{eqn:qk}
\end{align}
\end{itemize}
This completes the $k$th time step, yielding the sample $(l_k, a_k,\Tilde{y}_{k+1}, y_{k+1})$, along with the quantities required for the next step. We formalize this claim as Proposition \ref{prop}. 
\begin{algorithm}
\caption{Simulation algorithm for an individual $i$ up to the $K$th follow-up}\label{alg:sim2}
\begin{algorithmic}
    \State $k \leftarrow 0$
    \While {$k \leq K$}
        \State Sample $w_{k,1},w_{k,2}, w_{k,3}$ independent standard uniform variables
        \State $l_k \leftarrow F^{-1}_{L_k\mid Z,\bar{L}_{k-1},\bar{A}_{k-1},Y_k = 0} (w_{k,1} \cmid z,\bar {l}_{k-1},\bar {a}_{k-1}) $
        \State $a_k \leftarrow F^{-1}_{A_k \mid Z,\bar{L}_k,\bar{A}_{k-1},Y_k = 0}(w_{k,2}\cmid z,\bar {l}_k,\bar {a}_{k-1})$
        \If{k = 0}
            \State $v_{k,j} = w_{0,1}$
        \Else{}
            \For{$j \leftarrow 1, \ldots,k$}
                    \Comment{$L_j$ dist. in survivors per (\ref{eqn:ljyk})}
                \State $v_{k,j} \leftarrow \frac{v_{k-1,j} - C(v_{k-1,j},q_{k-1,j};\Theta_{k,k-j})} {1-q_{k-1,j}}$
            \EndFor
        \EndIf
        \State $\xi_{k+1} \leftarrow w_{k,3}$
         \For{$j\leftarrow 0,\ldots,k$}
            \Comment{`Unweaving' as per (\ref{eqn:ylk})}
            \State $\xi_{k-j} \leftarrow h^{-1}(\xi_{k+1-j},v_{k,k-j};\Theta_{K-j})$
        \EndFor
        \State $\Tilde{y}_{k+1} \leftarrow F_{\Tilde{Y}_{k+1}\mid \bar{A}_k, Y_{k-1} = 0}^{-1}(\xi_{0}\cmid do(\bar{a}_k))$
        \State $y_{k+1} = I(\Tilde{y}_{k+1} >1)$
        \If {$y_{k+1} = 1$}
            \State \textbf{Stop}
        \Else{}
        \Comment{Prepare $q$s for the next time step per (\ref{eqn:qk})}
        \State $q_{k,0} \leftarrow F_{\Tilde{Y}_{k+1}\mid \bar{A}_k, Y_{k-1} = 0}(\Tilde{Y}_{k+1} = 1\cmid do(\bar{a}_k))$
        \For{$j \leftarrow 1, \ldots k$}
            \State $q_{k,j} \leftarrow h\big \{ q_{k,j-1},v_{k,j-1} ; \Theta_{k,j-1}\big \}$
        \EndFor
        \State $k \leftarrow k+1$
        \EndIf
        \EndWhile
    \State \Return $(\bar{l}_k,\bar{a}_k,\Tilde{y}_{k+1},\bar{y}_{k+1})$
\end{algorithmic}
\end{algorithm}

Crucially, our method avoids computing the full conditional distribution of $\Tilde{Y}_{k+1}\mid \bar{L}_{k}, do(\bar{A}_k), Y_k = 0$, but rather requiring just a quantile corresponding to our simulated value; this is central to the efficiency of our approach.
(\ref{eqn:ljyk}) involves $k$ calculations of the distributions of each $L_{j} \cmid \bar{L}_{j-1}$ among those who survived to time $k$, in which all the quantities, including quantiles and copulas,  are available from the previous step $k-1$. In (\ref{eqn:ylk}), we recursively apply $k+1$ inverse $h$-functions to obtain the marginal quantile of $F(\Tilde{Y}_{k+1}\cmid do(\bar{a}_k),Y_{k} = 0)$. Finally, we simulate $\Tilde{Y}_{k+1}$ from the target marginal structural model via inversion. Before proceeding to the next time step $k+1$, we prepare the $q_{k,j}$ through $k + 1$ recursive computations in (\ref{eqn:qk}). In total, this time step requires $3k + 5$ functions, including the simulation of $L_k$ and $A_k$.

We proceed to show that the proposed method works using mathematical induction.
        
\begin{proposition}
Given the quantiles $w_{k,1}, w_{k,2}, w_{k,3}$ for each $k=0,\ldots,K-1$ and associated univariate distributions, together with copula parameters $\Theta_{k,j}$ for 
each $k=0,\ldots,K$ and $j=0,\ldots,k$, we can use the procedure outlined above to simulate a survival outcome according to the specified distributions.
\end{proposition}\label{prop}

The proof is given in Appendix \ref{sec:app_induction}.

In the derivation and proof presented above, we define the latent variable $\Tilde{Y}$ as the incremental time to failure which naturally aligns with the Cox marginal structural model that we use as a working example. Importantly, a key strength of our approach lies in its generality: our proposed algorithm is fully agnostic to the choice of survival model and holds for any specification of the latent variable and its distribution. This is a fundamental result that enables the direct simulation of survival outcomes from any causal marginal structural model, whether continuous or discrete.

\begin{remark}

    To illustrate this point, consider the case of simulating a binary survival outcome from a discrete marginal structural model, such as that defined in (\ref{eqn:disc_cox}). Similar to the framework in \cite{fan2017high}, we assume that there exists a continuous latent random variable, $S_{k+1}$, whose distribution under the intervention $do(\bar{A}_k)$ is $F_{S_{k+1} \mid do(\bar{A}_k),Y_k = 0}$, such that the binary survival indicator is obtained by dichotomizing $S_{k+1}$ at a threshold $C$. This formulation allows dependencies between $Y_{k+1}$ and past covariates to be modelled via the latent variable $S_{k+1}$, and the simulation procedure follows the algorithm described above, replacing $\Tilde{Y}_{k+1}$ by $S_{k+1}$. Given the marginal quantile $ u_{k+1}$, inversion yields $s_{k+1} = F^{-1}_{S_{k+1} \mid do(\bar{A}_k),Y_k = 0}(u_{k+1} \cmid do(\bar{a}_k),y_k = 0)$ Finally, the binary outcome is obtained as $y_{k+1} = I(s_{k+1} >C)$.
    

\end{remark}
In this basic setting, simulating $K$ time steps for each individual requires $\sum_{k=0}^{K-1}(3k+5) = 3K^2/2 + 7K/2$ calculations.This implies that the computational complexity scales quadratically with $K$. 

If additional time-varying confounders---either observed or unobserved---are incorporated, such that $\bs L$ becomes a $p$-dimensional vector, the number of iterations needed of (\ref{eqn:ljyk}), (\ref{eqn:ylk}) and (\ref{eqn:qk}) scales proportionally with $p$. As a result, the total number of computations becomes $\sum_{k=0}^{K-1}(3pk + 2p +3) = K(3K+1)p/2+3K$. Thus, computational complexity grows linearly with the number of confounders $p$. In the example presented in Appendix \ref{sec:app_seaman}, we replicate the simulation study by \cite{seaman2023simulating}, simulating data featuring two time-varying confounders. 

However, it should be noted that if an individual experiences failure at any time step, the simulation terminates early. Hence, these stated complexities represent the maximum possible computational cost per individual.
All of these calculations require evaluations of $h$-functions and their inverses and Copula distribution functions. For commonly used copulas, such evaluations are computationally efficient.

\subsection{Illustration: Concomitant Treatment} \label{sec:soul}


We illustrate how the proposed simulation framework can support sample size calculation under complex, real-world scenarios. To do so, we construct a synthetic study inspired by the SOUL trial (\textbf{S}emaglutide cardi\textbf{O}vascular o\textbf{U}tcomes tria\textbf{L}) \citep{mcguire2023effects}. The SOUL trial is a randomized controlled study assessing the effects of oral semaglutide, a
 glucagon-like peptide-1 receptor agonist (GLP-RA), on cardiovascular risk in individuals with type 2 diabetes, and cardiovascular and/or kidney diseases. The primary endpoint is time to first occurrence of major adverse cardiovascular events (MACE), analyzed under the intention-to-treat principle.

 A design challenge in SOUL arises from the permitted use of concomitant medications such as SGLT2 inhibitors (SGLT2-i), and there is demonstrably more uptake among participants in the control arm. The uneven `drop-in' treatment between the semaglutide and placebo arms may dilute the observed intention-to-treat effect, because both groups end up receiving some active treatment, blurring the distinction between them, and potentially undermining statistical power if unaccounted for during sample size calculation. Our goal is to show how the proposed simulation approach can capture such intercurrent events while preserving a predefined causal estimand.
 

We simulate data over seven half-yearly follow-up periods. Let $B\in \{0,1\}$ indicate baseline cardiovascular disease, $L_k$ denote HbA1c at time $k$, $A\in \{0,1\}$ the randomized treatment (semaglutide), $S_k\in \{0,1\}$ SGLT2-i use at time $k$ and $Y_k$ the binary survival status at time $k$. We assume perfect adherence to $A$ and model event times using a Cox marginal structural model:
\begin{align}
    \lambda(\Tilde{y}_{k+1}\cmid do(A,\bar{S}_k),Y_{k} = 0  ) = \lambda_0\exp\left(\beta_A \,A + \beta_S\,S_k + \beta_{AS}\,AS_k\right), \label{eqn:soul_msm}
\end{align}
where $\lambda_0$ is a baseline hazard assumed to be constant between follow-up visits and the survival indicator is calculated as $Y_{k+1} = I(\Tilde{Y}_{k+1} < 1)$. Baseline characteristics and model parameters are loosely based on the design assumptions of the SOUL trial \citep{mcguire2023effects}, including an annual MACE rate of 3.5\% and a hazard ratio of 0.83 for semaglutide (i.e.~$\beta_A = \log(0.83)$). We construct 10 simulation scenarios by varying both the prevalence (0\%, 5\%, 10\% and 15\% annual drop-in) in the placebo group and efficacy (hazard ratios of 0.95,0.9,0.85) of SGLT2-i use. These hazard ratios reflect estimates reported for SGLT2 inhibitors in \citet{mcguire2021association}. Full specifications of the data-generating model and scenario parameters are provided in Appendix \ref{sec:app_soul}. For each scenario and trial sizes of $n = \{5,\!000,10,\!000,20,\!000\}$, we simulate 400 datasets and fit a Cox MSM using inverse probability of treatment weights with stabilized weights. Naïve unweighted estimators are also computed for comparison. For reference, the total randomized population in SOUL is 9,650.

\begin{figure}
    \centering
    \includegraphics[width=0.9\linewidth]{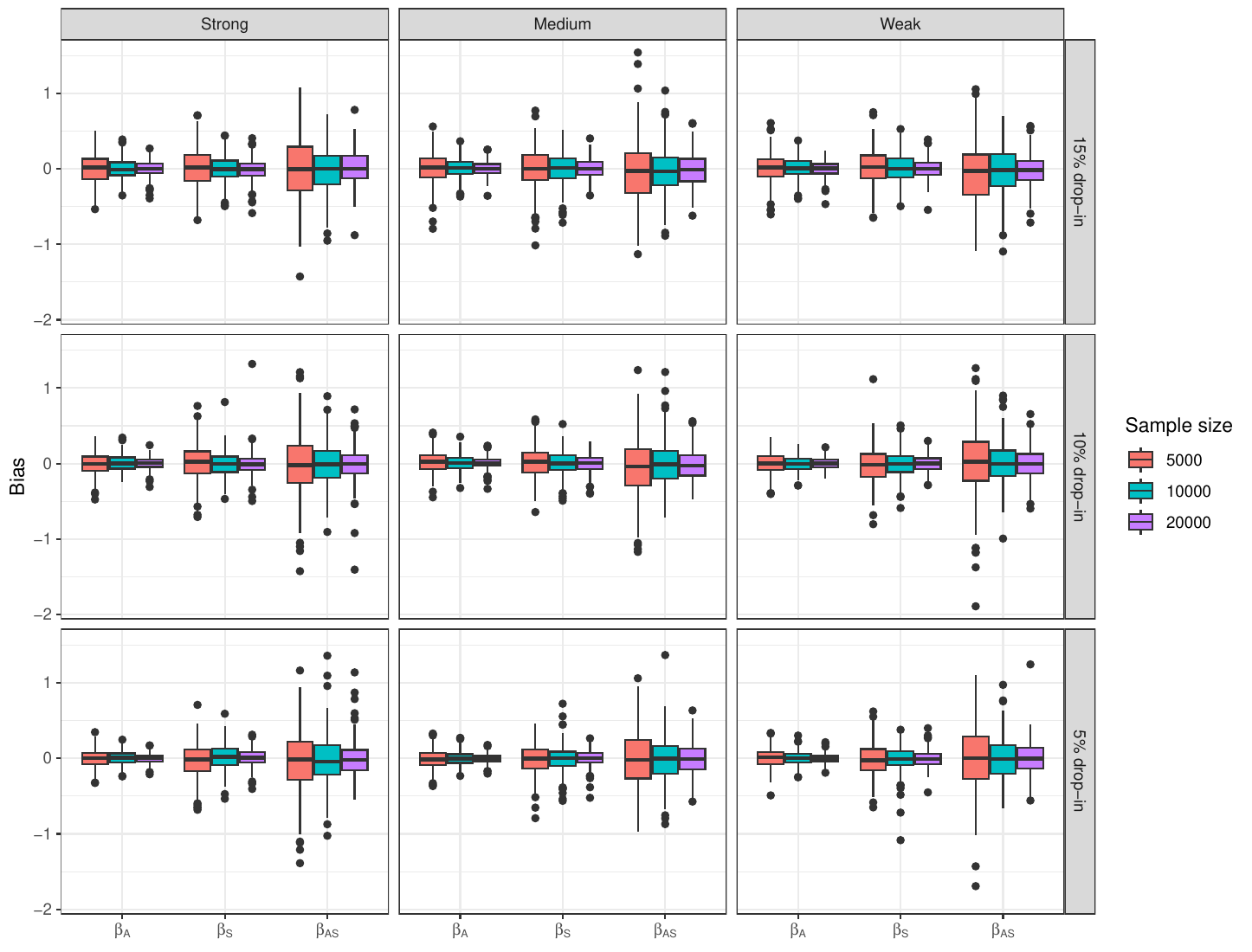}
    \caption{Boxplots of the bias for $\beta_A$, $\beta_S$ and $\beta_{AS}$ by inverse probability weighting (IPTW) across different drop-in scenarios.}
    \label{fig:bias_iptw}
\end{figure}

\begin{figure}
    \centering
    \includegraphics[width=0.8\linewidth]{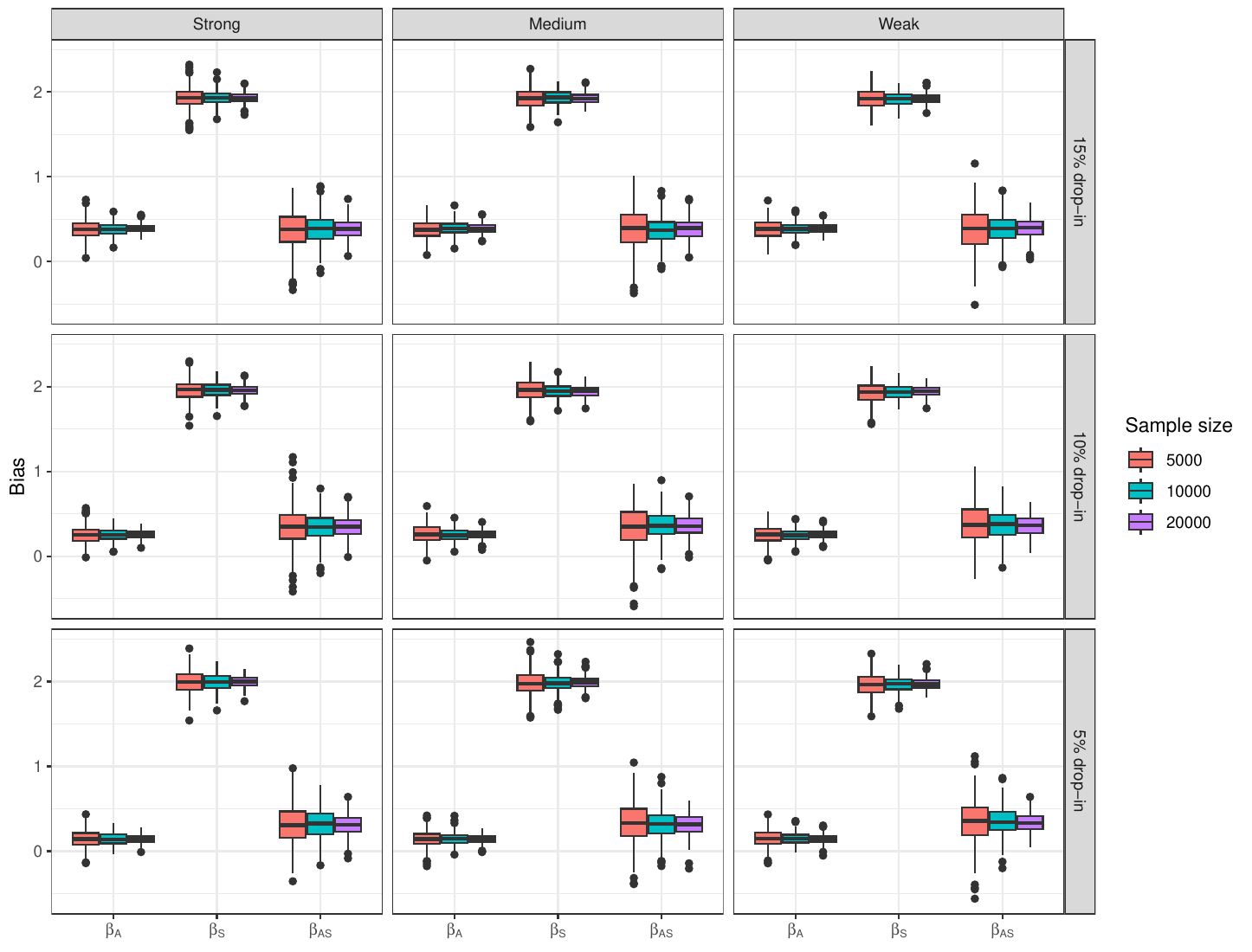}
    \caption{Boxplots of the bias for $\beta_A$, $\beta_S$ and $\beta_{AS}$ by unweighted Cox regression across different drop-in scenarios.}
    \label{fig:bias_unweighted}
\end{figure}

Figure \ref{fig:bias_iptw} and Figure \ref{fig:bias_unweighted} display boxplots of the estimation biases for the coefficients $\beta_A$, $\beta_S$ and $\beta_{AS}$ by inverse-probability-weighted and unweighted Cox regression, respectively. Across all scenarios, the inverse-probability-weighted estimates exhibit biases that are centered at zero. As the sample size increases, the distribution of these biases becomes narrower and more tightly concentrated around zero, indicating consistency. In contrast, unweighted estimates show substantial and persistent bias across all scenarios, with no evidence of convergence toward zero as the sample size increases. Table \ref{tab:converage} presents the empirical coverage: the inverse-probability-weighted method achieves coverage close to the nominal 95\% level, whereas the unweighted approach yields empirical coverage well below the nominal level, reflecting the impact of the time-varying confounding introduced to the data.

\begin{table}[ht]
\centering
\begin{adjustbox}{max width=\textwidth}
\begin{tabular}{llcccccccc}
\toprule
\multirow{2}{*}{\textbf{\makecell{Effect of \\SGLT2-i}}} & \multirow{2}{*}{\textbf{\makecell{Prevalence in \\ placebo}}} & \multirow{2}{*}{\textbf{\makecell{$n$}}} & \multicolumn{2}{c}{\textbf{$\beta_A$}} & \multicolumn{2}{c}{\textbf{$\beta_S$}} & \multicolumn{2}{c}{\textbf{$\beta_{AS}$}} \\ 
\cmidrule(lr){4-5} \cmidrule(lr){6-7} \cmidrule(lr){8-9}
& & & IPTW & Unweighted & IPTW & Unweighted & IPTW & Unweighted \\ 
\midrule
\textbf{Strong effect }&&$5000$ & $96.5\%$ & $65.0\%$ & $94.0\%$ & $0.0\%$ & $93.0\%$ & $74.0\%$ &  \\
&5\% drop-in&$10000$ & $94.8\%$ & $47.3\%$ & $95.0\%$ & $0.0\%$ & $94.3\%$ & $49.0\%$ &  \\
&&$20000$ & $95.0\%$ & $18.5\%$ & $95.0\%$ & $0.0\%$ & $95.0\%$ & $25.8\%$ &  \\

\cmidrule{2-9}

&&$5000$ & $94.0\%$ & $28.0\%$ & $95.0\%$ & $0.0\%$ & $92.3\%$ & $64.5\%$ &  \\
&10\% drop-in&$10000$ & $95.8\%$ & $4.5\%$ & $96.5\%$ & $0.0\%$ & $95.3\%$ & $42.5\%$ &  \\
&&$20000$ & $96.5\%$ & $0.0\%$ & $95.8\%$ & $0.0\%$ & $96.3\%$ & $16.3\%$ &  \\

\cmidrule{2-9}

&&$5000$ & $92.8\%$ & $4.8\%$ & $94.0\%$ & $0.0\%$ & $94.0\%$ & $60.8\%$ &  \\
&15\% drop-in&$10000$ & $94.5\%$ & $0.0\%$ & $96.0\%$ & $0.0\%$ & $94.3\%$ & $34.0\%$ &  \\
&&$20000$ & $93.0\%$ & $0.0\%$ & $95.0\%$ & $0.0\%$ & $94.0\%$ & $7.5\%$ &  \\

\midrule

\textbf{Moderate effect}&&$5000$ & $95.3\%$ & $68.0\%$ & $96.5\%$ & $0.0\%$ & $95.5\%$ & $69.3\%$ &  \\
&5\% drop-in&$10000$ & $94.5\%$ & $43.3\%$ & $93.8\%$ & $0.0\%$ & $94.3\%$ & $50.8\%$ &  \\
&&$20000$ & $95.3\%$ & $17.5\%$ & $96.0\%$ & $0.0\%$ & $93.8\%$ & $25.0\%$ &  \\

\cmidrule{2-9}

&&$5000$ & $96.0\%$ & $25.5\%$ & $97.3\%$ & $0.0\%$ & $95.5\%$ & $63.8\%$ &  \\
&10\% drop-in&$10000$ & $95.0\%$ & $4.5\%$ & $96.3\%$ & $0.0\%$ & $93.5\%$ & $40.8\%$ &  \\
&&$20000$ & $95.8\%$ & $0.3\%$ & $96.3\%$ & $0.0\%$ & $95.3\%$ & $14.8\%$ &  \\

\cmidrule{2-9}

&&$5000$ & $94.3\%$ & $4.8\%$ & $95.5\%$ & $0.0\%$ & $95.8\%$ & $58.0\%$ &  \\
&15\% drop-in&$10000$ & $95.3\%$ & $0.0\%$ & $94.5\%$ & $0.0\%$ & $94.0\%$ & $37.0\%$ &  \\
&&$20000$ & $92.0\%$ & $0.0\%$ & $95.3\%$ & $0.0\%$ & $94.0\%$ & $10.0\%$ &  \\

\midrule

\textbf{Weak effect}&&$5000$ & $96.3\%$ & $65.5\%$ & $95.3\%$ & $0.0\%$ & $93.5\%$ & $63.5\%$ &  \\
&5\% drop-in&$10000$ & $96.0\%$ & $42.8\%$ & $97.0\%$ & $0.0\%$ & $95.3\%$ & $45.3\%$ &  \\
&&$20000$ & $95.0\%$ & $13.3\%$ & $95.0\%$ & $0.0\%$ & $95.0\%$ & $18.5\%$ &  \\

\cmidrule{2-9}

&&$5000$ & $96.0\%$ & $26.8\%$ & $96.0\%$ & $0.0\%$ & $94.5\%$ & $59.8\%$ &  \\
&10\% drop-in&$10000$ & $96.3\%$ & $5.0\%$ & $96.5\%$ & $0.0\%$ & $95.0\%$ & $37.0\%$ &  \\
&&$20000$ & $95.5\%$ & $0.0\%$ & $96.8\%$ & $0.0\%$ & $96.3\%$ & $11.5\%$ &  \\

\cmidrule{2-9}

&&$5000$ & $93.3\%$ & $5.3\%$ & $95.8\%$ & $0.0\%$ & $93.8\%$ & $57.5\%$ &  \\
&15\% drop-in&$10000$ & $93.3\%$ & $0.0\%$ & $93.5\%$ & $0.0\%$ & $92.5\%$ & $34.8\%$ &  \\
&&$20000$ & $95.5\%$ & $0.0\%$ & $95.5\%$ & $0.0\%$ & $95.0\%$ & $7.3\%$ &  \\
\midrule
&&$5000$ & $96.0\%$ & $96.0\%$ & $-$ & $-$ & $-$ & $-$ &  \\
&No drop-in&$10000$ & $94.5\%$ & $94.5\%$ & $-$ & $-$ & $-$ & $-$ &  \\
&&$20000$ & $93.5\%$ & $93.5\%$ & $-$ & $-$ & $-$ & $-$ &  \\

\bottomrule
\end{tabular}
\end{adjustbox}
\caption{Empirical coverage rates of 95\% confidence intervals of inverse-probability-weighted (IPTW) and unweighted Cox regression methods.}
\label{tab:converage}
\end{table}

These results confirm that the simulated data adhere to the marginal structural model specified in (\ref{eqn:soul_msm}), and that time-varying confounding is induced. We now demonstrate how the proposed simulation framework can inform trial design, specifically for sample size determination under realistic scenarios involving intercurrent events.

 The SOUL trial was designed to evaluate the effect of semaglutide on CV outcomes under the intention-to-treat principle, contrasting the outcomes of individuals randomized to semaglutide ($A = 1$) versus placebo ($A = 0$). Accordingly, we fit the following Cox proportional hazards model:
\begin{align*}
    \lambda(t\cmid A) = \lambda_0\exp(\beta_{ITT} A),
\end{align*}
where $t$ is time to the first occurrence of MACE and $\exp(\beta_{ITT})$ represents the hazard ratio associated with being assigned semaglutide, irrespective of SGLT2-i use. Table~\ref{tab:itt_power} reports the estimated intention-to-treat effects and empirical power based on a one-sided type I error rate of $0.025$. In the hypothetical scenario with no SGLT2-i use in either arm, the estimated intention-to-treat effect closely approximates the true hazard ratio for semaglutide (0.83). Under this scenario, a trial with 10,000 participants achieves approximately 85\% power. However, in all other scenarios, the intention-to-treat estimates are attenuated due to drop-in and consequently reduce power. To maintain at least 80\% power, our simulations suggest that a sample size of 20,000 is sufficient but may be unnecessarily large and resource-intensive. Conversely, a trial with 5,000 participants is consistently underpowered, even in the absence of drop-in. A sample size of 10,000 appears to strike a reasonable balance: it achieves adequate power in scenarios with weak SGLT2-i effects, but falls short when the effect of SGLT2-i is strong or if the prevalence of drop-in in the placebo arm exceeds 5\%. For the moderate scenarios, where SGLT2 inhibitors have a moderate effect, 10,000 participants is borderline sufficient, but if drop-in rates in the placebo arm exceed 5\%, power may fall below acceptable thresholds. These findings suggest that 10,000 participants may be an appropriate design choice, but the sample size should be adjusted based on the plausibility of drop-in and effect size assumptions.


The 10 scenarios considered here are illustrative. The proposed simulation framework is highly flexible. For example, users can vary the marginal effect of semaglutide by adjusting $\beta_A$, alter the baseline incidence rate, or introduce more time-varying covariates. Because the algorithm relies on analytic function evaluations, it is computationally efficient: simulating 10,000 individuals over seven time steps requires approximately 40 seconds on a single core. This efficiency makes it practical and feasible to conduct large-scale simulation studies using parallel computation to support power analyses and inform robust trial design.

\begin{table}[t]
\centering
\begin{adjustbox}{max width= 0.7\textwidth}
\begin{tabular}{llcccc}
\toprule
\textbf{Effect of SGLT2-i} & \textbf{\makecell{Prevalence in \\ placebo}} & \textbf{n} & \textbf{Estimate} & \textbf{\makecell{Hazard \\ Ratio}}  & \textbf{Power}\\ 
\midrule
\textbf{Strong effect }&&$5000$ & $-0.164$ & $0.849\%$ & $46.5\%$ \\
&5\% drop-in&$10000$ & $-0.165$ & $0.848\%$ & $74.3\%$ \\
&&$20000$ & $-0.169$ & $0.845\%$ & $97.5\%$ \\
\cmidrule{2-6}
&&$5000$ & $-0.153$ & $0.858\%$ & $43.0\%$ \\
&10\% drop-in&$10000$ & $-0.148$ & $0.862\%$ & $65.5\%$ \\
&&$20000$ & $-0.151$ & $0.860\%$ & $93.0\%$ \\
\cmidrule{2-6}
&&$5000$ & $-0.143$ & $0.867\%$ & $37.5\%$ \\
&15\% drop-in&$10000$ & $-0.143$ & $0.867\%$ & $63.8\%$ \\
&&$20000$ & $-0.135$ & $0.874\%$ & $87.8\%$ \\
\midrule
\textbf{Moderate effect}&&$5000$ & $-0.176$ & $0.839\%$ & $52.0\%$ \\
&5\% drop-in&$10000$ & $-0.173$ & $0.841\%$ & $81.0\%$ \\
&&$20000$ & $-0.179$ & $0.836\%$ & $98.8\%$ \\

\cmidrule{2-6}

&&$5000$ & $-0.160$ & $0.852\%$ & $45.0\%$ \\
&10\% drop-in&$10000$ & $-0.164$ & $0.848\%$ & $75.8\%$ \\
&&$20000$ & $-0.161$ & $0.851\%$ & $97.8\%$ \\
\cmidrule{2-6}

&&$5000$ & $-0.161$ & $0.851\%$ & $46.0\%$ \\
&15\% drop-in&$10000$ & $-0.157$ & $0.855\%$ & $76.3\%$ \\
&&$20000$ & $-0.154$ & $0.857\%$ & $94.5\%$ \\
\midrule
\textbf{Weak effect}&&$5000$ & $-0.174$ & $0.841\%$ & $52.0\%$ \\
&5\% drop-in&$10000$ & $-0.181$ & $0.834\%$ & $84.5\%$ \\
&&$20000$ & $-0.184$ & $0.832\%$ & $98.3\%$ \\
\cmidrule{2-6}
&&$5000$ & $-0.176$ & $0.838\%$ & $53.3\%$ \\
&10\% drop-in&$10000$ & $-0.180$ & $0.835\%$ & $82.8\%$ \\
&&$20000$ & $-0.176$ & $0.839\%$ & $98.3\%$ \\
\cmidrule{2-6}
&&$5000$ & $-0.182$ & $0.834\%$ & $55.0\%$ \\
&15\% drop-in&$10000$ & $-0.176$ & $0.838\%$ & $82.3\%$ \\
&&$20000$ & $-0.173$ & $0.841\%$ & $98.3\%$ \\
\midrule
&&$5000$ & $-0.187$ & $0.830\%$ & $56.8\%$ \\
&No drop-in&$10000$ & $-0.183$ & $0.833\%$ & $85.0\%$ \\
&&$20000$ & $-0.187$ & $0.830\%$ & $99.5\%$ \\

\bottomrule
\end{tabular}
\end{adjustbox}
\caption{Under the intention-to-treat principal, estimates of $\beta_{ITT}$, hazard ratios ($\exp{\beta_{ITT}}$), and power using a one-sided type I error rate of $0.025$ for various scenarios and sample sizes ($n$).}\label{tab:itt_power}
\end{table}

\subsection{Competing risks} \label{sec:comp_risk}
An important consideration in research concerning time-to-event outcomes is that of competing risks; these are events whose occurrence prevents the observation of the outcome of interest---for example, non-cardiovascular death in studies of MACE. Classical literature defines various statistical estimands for competing risk settings, including the marginal, cause-specific and subdistribution hazards \citep[see, for example,][]{pintilie2007analysing, geskus2016data,young2020causal}. These estimands differ in interpretation and causal relevance.

Our proposed simulation framework offers the flexibility to  efficiently simulate data from marginal structural models that target the \textit{controlled direct effect}:
\begin{align}
    F(\Tilde{Y}_{k+1} \cmid do(\bar{A}_k = \bar{a}_k,\bar{D}_k = \bar{0}_k ), Z = z, Y_{k} = 0) = f(\bar{a}_k,z; \alpha), \label{eqn:comp_msm1}
\end{align}
where $\{D_k\}_{k\in \mathcal{K}_+}$ denote the sequence of competing events and $do(\bar{D}_k = \bar{0}_k )$ represents a hypothetical intervention eliminating these events. This aligns with the definition of marginal hazard in \cite{young2020causal}. Under this framework, competing risks are essentially treated as equivalent to censoring. To generate the competing event process, two approaches can be considered. The first specifies a fully conditional distribution: $F(\Tilde{D}_{k+1} \cmid \bar{L}_k,\bar{A}_k, Z, Y_k = 0) $. Under this approach, the outcome process $Y_{1:K}$ is simulated according to Algorithm \ref{alg:sim2}, targeting the causal margin defined in (\ref{eqn:comp_msm1}).  Once simulation of the failure process is complete---suppose the trajectory terminates at time step $K'$  when $Y_{K'+1} = 1$---the competing event process is then simulated for each $k \in \{0,\ldots, K'\}$. At each of these time steps, $\Tilde{D}_{k+1}$ is drawn from the specified conditional distribution $F(\Tilde{D}_{k+1} \cmid \bar{L}_k,\bar{A}_k, Z, Y_k = 0) $. If a competing event occurs before time $K'$, the individual's trajectory is truncated at the corresponding time step. Censoring events, such as loss to follow-up, can also be simulated using this approach.

Alternatively,  a marginal structural model may be specified for the competing event hazard,  mirroring the approach used for the primary outcome:
\begin{align}
F(\Tilde{D}_{k+1} \cmid do(\bar{A}_k = \bar{a}_k,\bar{Y}_k = \bar{0}_k ), Z = z, D_{k} = 0) = g(\bar{a}_k,z; \beta)  \label{eqn:comp_msm2}.
\end{align}
Additionally, we impose the following conditional independence: 
$$
\Tilde{D}_{k+1} \indep \Tilde{Y}_{k+1} \cmid do(\bar A_k, \bar{Y}_k = \bar{0}_k, \bar{D}_k = \bar{0}_k), Z, \bar{L}_k,
$$
which states that, under intervention, and conditional on baseline and time-varying covariates, $\Tilde{D}_{k+1}$ and $\Tilde{Y}_{k+1}$ are independent. Graphically, this corresponds to the absence of a direct edge between $Y_{k+1}$ and $D_{k+1}$ in the underlying causal graph---an edge that would not be identifiable from observed data in any case.

 If both the outcome and the competing event are simulated using MSMs, the simulation proceeds as follows:
\begin{enumerate}
    \item Simulate baseline covariates $Z$. Set $Y_0=D_0=0$.
    \item For time point $k = 0,\dots,K$:
    \begin{enumerate}
        \item If $Y_k = 0$ or $D_k = 0$, simulate $L_k, A_k$; otherwise, stop;
        \item If $Y_k = 0$, simulate $Y_k$ following Algorithm \ref{alg:sim2}; otherwise set $Y_k$ to null;
        \item If $D_k = 0$, simulate $D_k$ following Algorithm \ref{alg:sim2}, substituting $Y$ with $D$, along with the corresponding distributions and copulas; otherwise, set $D_k$ to null.
    \end{enumerate}
    \item Truncate this individual's trajectory at the first occurrence of either the outcome of interest or the competing event.
\end{enumerate}

Notably, in step (b), the decision to simulate $Y_{k+1}$ depends solely on $Y_{k}$ irrespective of $D_{k}$ and vice versa for step (c). This reflects the fact that we target a \textit{controlled direct effect}, as specified in (\ref{eqn:comp_msm1}) and (\ref{eqn:comp_msm2}), where the other event is conceptually eliminated through intervention. 


This approach is particularly helpful in settings like the estrogen therapy example in \citet{young2020causal}. By explicitly specifying the direction and magnitude of the direct effects of treatment on both prostate cancer mortality and other-cause mortality, one can simulate data to understand how these effects jointly determine the total effect, i.e.~the cause-specific hazard. This framework can inform trial design: clinicians can input plausible direct effect assumptions and compute the sample size required to achieve adequate power for detecting a treatment effect on the cause-specific hazard.

\section{Discussion}\label{sec:discuss}

In this paper, we proposed and demonstrated a principled strategy for simulating longitudinal data exactly from marginal structural models, leveraging the frugal parameterization \citep{Evans2024}. Our proposed method is highly flexible and can accommodate a variety of practical scenarios, including multiple treatments (for example, primary and drop-in), dynamic treatment regimes, dose-response analyses, multiple outcomes (such as secondary outcomes and competing risks), and censoring. Additionally, it imposes no restrictions on the data-generating process or marginal structural causal models.
In Section \ref{sec:survival}, we extended this method to survival models and time-to-event outcomes, which are common in clinical trials. 
The method’s computational efficiency, combined with its analytic precision, makes it well-suited to practical tasks such as simulation-based method evaluation and sample size determination.

In Section \ref{sec:comp_risk}, we discussed how our proposed simulation framework can flexibly generate data with competing risks from marginal structural models targeting the marginal hazard. However, extending the framework to simulate data under alternative estimands---specifically, the cause-specific and subdistribution hazards, which represent total effects---remains nontrivial. Future work may explore how to extend the proposed method to support simulation from marginal structural models that target these alternative causal quantities.

A limitation of the method is difficulty in enforcing Markov-type conditional independences between the outcome and the history of time-varying confounders. For instance, the pair-copula construction we employed in Section \ref{sec:frugal_y} to model the association between $Y$ and $L_0,\dots,L_k$ makes it difficult to impose restrictions that might be considered natural; for example, that the distribution of the outcome only depends upon the most recent few values of a time-varying covariate. Parameterizing such a model is infeasible unless all the bivariate copulas in the pair-copula construction are Gaussian; in this case, conditional independence can be enforced by appropriately choosing the partial correlation parameters. 
If enforcing such Markov conditions is absolutely necessary, we would suggest using either this Gaussian approximation or numerical integration. 
However, numerical integration in high-dimensions is well known to be infeasible.


\appendix

\section{Extension to include multiple time-varying variables}\label{sec:app_latent}
We outline how the proposed algorithm in Section \ref{sec:method} can be extended to include multiple time-varying covariates. From a simulation perspective, there is no distinction between whether such covariate is observed or unobserved; for illustrative purposes, we assume that the additional time-varying variable $H$ follows a latent process$\{H_k\}_{k \in \mathcal{K}}$. The steps for simulate data from the model depicted in Figure \ref{fig:longi_latent} are as follows:
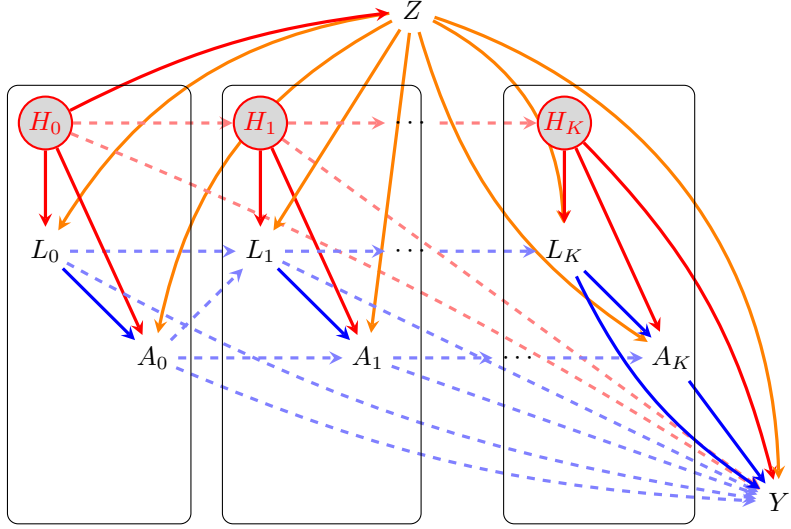
\begin{figure}
  \begin{center}
  \begin{tikzpicture}
  [rv2/.style={rv, inner sep=0.1mm}, node distance=20mm, >=stealth]
  \usetikzlibrary{backgrounds} 
  \pgfsetarrows{latex-latex};
\begin{scope}
    \node[rvs0] (Z0) {$\L_0$};
    \node[lv, above of=Z0, yshift = -3mm] (U0) {$H_0$};
    \node[rvs0, below right of=Z0] (X0) {$\A_0$};
    \draw[rounded corners] ($(U0)+(-5mm,+5mm)$) rectangle ($(X0)+(5mm,-22mm)$);
    \node[rvs0, above right of=X0] (Z1) {$\L_1$};
    \node[lv, above  of=Z1, yshift = -3mm] (U1) {$H_1$};
      \draw[rounded corners] ($(U1)+(-5mm,+5mm)$) rectangle ($(X1)+(19mm,-22mm)$);
    \node[rvs0, below right of=Z1] (X1) {$\A_1$};
    \node[right of=U1] (d0) {$\cdots$};
    \node[right of=Z1] (d1) {$\cdots$};
    \node[right of=X1] (d2) {$\cdots$};
    \node[rvs0, above of=d0, yshift=-5mm] (C) {$Z$};
    \node[rvs0, right of=d1] (ZT) {$\L_K$};
    \node[lv, above of=ZT, , yshift = -3mm, inner sep=0mm] (UT) {$H_K$};
    
    \draw[rounded corners] ($(UT)+(-8mm,+5mm)$) rectangle ($(XT)+(21mm,-22mm)$);
    \node[rvs0, right of=d2] (XT) {$\A_K$};
    \node[rvs0, below right of=XT, xshift=0mm, yshift=-5mm] (YT) {$Y$};
    
    \begin{pgfonlayer}{background}
    \draw[deg, orange] (C) to[bend right=25] (Z0);
    \draw[deg, orange] (C) to[bend right=25] (X0);
    \draw[degl] (U0) -- (Z0);
    \draw[deg,dashed, blue!50] (Z0) -- (Z1);
    \draw[deg] (Z0) -- (X0);
    \draw[deg,dashed, blue!50] (X0) -- (Z1);
    \draw[degl] (U0) to[bend left=10] (C);
    \draw[degl] (U0) to (X0);
    \draw[degl,dashed, red!50] (U0) -- (U1);
    \draw[degl,dashed, red!50] (U0) to[bend left=5] (YT);
    \draw[degl,dashed, red!50] (U1) -- (YT);
    \draw[degl] (U1) -- (Z1);
    \draw[degl] (U1) to (X1);
    \draw[deg] (Z1) -- (X1);
    \draw[deg, orange] (C) -- (X1);
    \draw[deg,dashed, blue!50] (X0) -- (X1);
    \draw[deg, orange] (C) to[bend right=20] (XT);
    \draw[deg, orange] (C) -- (Z1);
    \draw[deg, orange] (C) to[bend left] (ZT);
    \draw[degl,dashed, red!50] (U1) -- (d0);
    \draw[deg,dashed, blue!50] (Z1) -- (d1);
    \draw[deg,dashed, blue!50] (X1) -- (d2);
    \draw[degl,dashed, red!50] (d0) -- (UT);
    \draw[deg,dashed, blue!50] (d1) -- (ZT);
    \draw[deg,dashed, blue!50] (d2) -- (XT);
    \draw[deg] (ZT) -- (XT);   
    \draw[degl] (UT) -- (ZT);
    \draw[degl] (UT) -- (XT);
    \draw[degl] (UT) to[bend left=15] (YT);   
    \draw[deg, orange] (C) to[bend left=35] (YT);
    \draw[deg,dashed, blue!50] (Z1) to[bend right=0] (YT);
    \draw[deg,dashed, blue!50] (Z0) to[bend right=10] (YT);
    \draw[deg,dashed, blue!50] (X1) -- (YT);
    \draw[deg,dashed, blue!50] (X0) to[bend right=10] (YT);
    \draw[deg] (XT) -- (YT);
    \draw[deg] (ZT) to[bend right=15] (YT);
    \end{pgfonlayer}

  \end{scope}
  \end{tikzpicture}
 \caption{Extension of the model in Figure \ref{fig:longi} which allows for an unobserved process $H_k$ affecting only $\X$, $\L_k$ and $Y$.}
  \label{fig:longi_latent}
  \end{center}
\end{figure}

\begin{enumerate}
    \item \textbf{Baseline simulation}: The directed acyclic graph in Figure \ref{fig:longi_latent} indicates that the latent variable $H_0$ precedes $Z$ in topological order, potentially indicating an initial selection bias. Accordingly, we simulate baseline values $(z, h_0, l_0, a_0)$ sequentially:
    \begin{align*}
        h_0 &=  F_{H_0}^{-1}(w_{0,0})\\
        z &=  F_{Z\mid H_0}^{-1}(w_Z \cmid h_0)\\
        l_0 &= F_{L_0\mid ZH_0}^{-1}(w_{0,1}\cmid z, h_0)\\
        a_0 &= F_{A_0\mid ZH_0L_0}^{-1}(w_{0,2}\cmid z,h_0,l_0),
    \end{align*}
    where $w_Z, w_{0,0}, w_{0,1}, w_{0,2}$ are standard uniform variables.
    \item \textbf{Time-step simulation}: For each time step $k=1,\dots,K$, simulate:
    \begin{align*}
        h_k &= F_{H_k\mid \bar{H}_{k-1}}^{-1}(w_{k,1} \cmid \bar{h}_{k-1})\\
        l_k &= F_{L_k \mid Z\bar{H}_k\bar{L}_{k-1}\bar{A}_{k-1}}^{-1}(w_{k,2},\cmid z, \bar{h}_k,\bar{l}_{k-1}, \bar{a}_{k-1})\\
        a_k &= F_{A_k\mid Z\bar{H}_k\bar{L}_{k}\bar{A}_{k-1}}^{-1}(w_{k,3},\cmid z, \bar{h}_k, \bar{l}_{k}, \bar{a}_{k-1})
    \end{align*}
    where $w_{k,1},w_{k,2},w_{k,3} \sim U(0,1)$.
    \item \textbf{Outcome simulation}: Once all time-varying covariates and treatment have been simulated, we apply the same derivation following the frugal parameterization as in Section \ref{sec:frugal_y} and simulate $Y$ in two steps:
\begin{enumerate}
    \item Compute the marginal quantile:
    \begin{align*}
        u^*_Y &= h^{-1} \big [\dots h^{-1}[h^{-1} \left \{h^{-1}\left (w_y,F(l_K \cmid \bar{l}_{K-1}, \bar{h}_K, z, \bar{a}_{K-1}); \Theta_{K,1} \right ),F(h_{K} \cmid \bar{l}_{K-1}, \bar{h}_{K-1}, z, \bar{a}_{K-1}) ; \Theta_{K,2} \right \} \notag \\
        &  F(l_{K-1} \cmid \bar{l}_{K-2}, \bar{h}_{K-1}, z, \bar{a}_{K-2}); \Theta_{K-1,1}] \dots ; \Theta_{0,2} \big]
    \end{align*}
    where $w_{Y} \sim \operatorname{U}(0,1)$;
    \item Obtain the outcome:
    $y = F_{Y\mid Z,\bar{A}_K}^{*{-1}}(u^*_Y \cmid z, do(\bar{a}_K))$. 
\end{enumerate}
\end{enumerate}

Compared to the case where without latent variable, the number of applications of inverse $h$-functions in Step 3(a) doubles. These transformations systematically disentangle the dependencies between $Y$ and $H_k$, as well as between $Y$ and $L_k$. In addition to the information required in the case without latent confounders, the user also specifies:
\begin{itemize}
    \item the development of the latent confounder $H_k$ as influenced by $Z$ and treatment history $\bar{A}_k$;
    \item $K+1$ pair-copula densities describing dependencies between the outcome and each member of the latent process: 
    $c_{YH_k\mid Z, \bar{H}_{k-1}, \bar{L}_{k-1},do(\bar{A}_{k-1})}$ for $k=0,1,\ldots,K$.
\end{itemize}

Extending the framework to include additional time-varying covariates follows a similar approach. In essence, simulating a latent process is technically equivalent to simulating any other time-varying confounder. Whether a confounder is observed or unobserved does not alter the fundamental simulation methodology. With common choices of parametric bivariate copulas, evaluating the $h$-functions is fast, making the simulation efficient. As the dimensionality of time-varying confounders, either observed or unobserved, the computational complexity grows linearly.

\section{Proof of Lemma 5.1}\label{sec:app_lemma}
\begin{proof} 
By definition:
        \begin{align}
    \lefteqn{F(l_j \cmid \bar{l}_{j-1},do(\bar{a}_{k}), y_k = 0) }\nonumber\\
    &=  \frac{\int_{1}^{\infty} F(l_j \cmid \bar{l}_{j-1},do(\bar{a}_{k}), \Tilde{y}_k, y_{k-1} = 0) \cdot p(\Tilde{y}_k \cmid \bar{l}_{j-1},do(\bar{a}_{k}), y_{k-1} = 0) \,d\Tilde{y}_k}{\int_{1}^{\infty} p(\Tilde{y}_k \cmid \bar{l}_{j-1},do(\bar{a}_{k}),  y_{k-1} = 0) \, d\Tilde{y}_k} \nonumber\\
     &= \frac{\int_{q_{k-1,j}}^{1} F(l_j \cmid \bar{l}_{j-1},do(\bar{a}_{k}), \Tilde{y}_k, y_{k-1} = 0)  \,dF(\Tilde{y}_k \cmid \bar{l}_{j-1},do(\bar{a}_{k}), y_{k-1} = 0)} {1-q_{k-1,j}}, \label{xxx}
\end{align}
where $q_{k-1,j} = F(\Tilde{y}_k = 1 \cmid \bar{l}_{j-1},do(\bar{a}_{k}), y_{k-1} = 0)$. Recall the result from \cite{joe1996families} in (\ref{eqn:pcc_cdf}), let $v_{k-1,j} = F(l_j \cmid \bar{l}_{j-1},do(\bar{a}_{k}),y_{k-1} = 0)$ and $u=F(\Tilde{y}_k \cmid \bar{l}_{j-1},do(\bar{a}_{k}),  y_{k-1} = 0)$, then the numerator in (\ref{xxx}) can be written as:
\begin{align*}
\int_{q_{k-1,j}}^{1} C(v_{k-1,j}\cmid u) \, d u &= \int_{q_{k-1,j}}^{1}  \frac{\partial }{\partial u} C(v_{k-1j},u) \, d u\\
     & = C(v_{k-1j},1) - C(v_{k-1j},q_{k-1,j})\\
     & = v_{k-1j} - C(v_{k-1j},q_{k-1,j}) 
\end{align*}
where the copula $C$ is $C_{L_j\Tilde{Y}_k\mid \bar{L}_{j-1},do(\bar{A}_k),Y_{k-1} = 0}(\cdot ;\Theta_{k-1,j})$, specifically.
\end{proof}

\section{Proof of Proposition 5.2}\label{sec:app_induction}
\begin{proof}
We prove that the algorithm correctly simulates data for all $k \geq 0$.

\textbf{Base case ($k=0$):} 

Initialize the algorithm by specifying the cumulative distribution functions $F_{L_0}$, $F_{A_0\mid L_0}$, $F_{\Tilde{Y}_1\mid do(A_0)}$, and the copula $C_{\Tilde{Y}_1L_0\mid do(A_0)}$, parameterized by $\Theta_{0,0}$. Draw $w_{0,1}$, $w_{0,2}$, $w_{0,3} \sim U(0,1)$ and compute:
\begin{align*}
    l_0 = F^{-1}_{L_0}(w_{0,1}) && a_0 = F^{-1}_{A_0 \mid L_0}(w_{0,2} \cmid l_0), 
\end{align*}
to obtain a sample $(l_0, a_0)$. Using the transformation:
 \begin{align*}
    F_{\Tilde{Y}_1 \cmid do(A_0)}(\Tilde{y}_1 \cmid do(a_0)) &= h^{-1}\big\{F_{\Tilde{Y}_1\cmid L_0, do(A_0)}(\Tilde{y}_1\cmid l_0, do(a_0)),F_{L_0}(l_0); \Theta_{0,0} \big\}, 
\end{align*}
set $w_{0,3} = F_{\Tilde{Y}_1\cmid L_0, do(A_0)}(\Tilde{y}_1\cmid l_0, do(a_0))$ and substitute $F_{L_0}(l_0)$ with $w_{0,1}$, yielding:
\begin{align*}
    u_1 = h^{-1}\left (w_{0,3},w_{0,1}; \Theta_{0,0} \right). 
\end{align*}

By inversion, compute $\Tilde{y}_1 = F_{Y_1\mid do(A_0)}^{-1}(u_1)$, and the corresponding binary outcome $y_1 = I(\Tilde{y}_1 <1) $. This completes the first iteration, yielding a sample $(l_0, a_0,\Tilde{y}_1, y_1)$. If $y_1 = 1$, stop; otherwise, before proceeding to $k=1$, we store the following values:
\begin{align*}
    u_{0,0} = F_{L_0}(l_0) && q_{0,0} = F_{\Tilde{Y}_1  \mid do(A_0)}(\Tilde{y}_1 = 1 \cmid do(a_0)).
\end{align*}

For $k=1$, define the necessary cumulative distribution functions $F_{L_1 \cmid L_0, A_0}$, $F_{A_1\mid \bar{L}_1, A_0}$, $F_{\Tilde{Y}_2\mid do(\bar{A}_1)}$, and copulas $C_{\Tilde{Y}_2L_1\mid L_0 do(\bar{A}_1)}$ and $C_{\Tilde{Y}_2L_0\mid do(A_0)}$, parameterized by $\Theta_{1,1}$ and $\Theta_{1,0}$, respectively. These two copulas associate $\Tilde{Y}_2$ with previous time-varying confounders. Draw $w_{1,1}$, $w_{1,2}$, $w_{1,3}$ from $U(0,1)$, and obtain $(l_1,a_1)$ as before.

We now need to obtain a sample of $\Tilde{y}_2$. Calculate:
\begin{align}
    F(l_0\cmid do(a_0), y_1 = 0) &= \frac{u_{0,0} - C^*_{\Tilde{Y}_1 L_0 \mid \A_0}(u_{0,0}, q_{0,0}; \Theta_{0,0})}{1 - q_{0,0}} \label{eqn:proof_k=1L0}\\
    F(\Tilde{y}_2\cmid l_0, do(\bar{a}_1), y_1 = 0) &=  h^{-1}\big\{ F(\Tilde{y}_2\cmid \bar{l}_1,do(\bar{a}_1), y_1 =0), F(l_1\cmid l_0,do(\bar{a}_1), y_1 = 0); \Theta_{1,1}\big\} \nonumber\\ 
    F(\Tilde{y}_2 \cmid do(a_1),y_1 = 0) &= h^{-1}\big\{ F(\Tilde{y}_2\cmid l_0, do(\bar{a}_1), y_1 = 0) ,F(l_0 \cmid do(\bar{a}_1), y_1 = 0); \Theta_{1,0} \big\}. \nonumber 
\end{align}
Note that we omit some subscripts on quantiles for brevity. In (\ref{eqn:proof_k=1L0}) we derive the quantile of $L_0$ given $a_0$ among survivors the start of the $k=1$ interval, where $u_{0,0}$ and $q_{0,0}$ are already computed in step $k = 0$. All other necessary quantities are specified at $k=1$. These computations yields a sample $(l_1,a_1,\Tilde{y}_2,y_2)$. If $y_2 = 1$, the simulation stops; otherwise, we store and compute the following values:
\begin{align*}
    u_{1,0} &= F_{L_0 \mid Y_1}(l_0 \cmid y_1 = 0)\\
    q_{1,0} &= F_{\Tilde{Y}_2  \mid do(\bar{A}_1) Y_1}(\Tilde{y}_2 = 1 \cmid do(\bar{a}_1), y_1 = 0)\\
    u_{1,1} &= F_{L_1 \mid L_0 Y_1}(l_0 \cmid l_0, y_1 = 0) \\
    q_{1,1} &= F_{\Tilde{Y}_2  \mid L_0 do(\bar{A}_1) Y_1}(\Tilde{y}_2 = 1 \cmid l_0,do(\bar{a}_1), y_1 = 0)\\
    & = h\big \{ q_{1,0}; u_{1,0}; \Theta_{1,0} \big \},
\end{align*}
where we simply store the first three quantities, which are directly available from the current step, while only the last one requires computation. Therefore, the base case is satisfied since the algorithm simulates the first time step and provides the required inputs for the next.

\textbf{Inductive step:} 

Assume the algorithm correctly simulates up to the $k$th iteration. We have $F(l_k \mid \overline{l}_{k-1}, do(\bar{a}_{k-1}), y_k=0)$ and
$F(y_{k+1} \mid \overline{l}_k, do(\bar{a}_{k}), y_k=0)$, as well as a collection of pair copulas between 
$Y_{k+1}$ and $L_j$  given $(\bar{l}_{j-1}, \bar{a}_{j-1})$ with parameters $\Theta_{k,j}$.  We need to compute the updated quantiles 
$F(l_j \mid \overline{l}_{j-1}, do(\bar{a}_{k}),y_k=0)$
for each $j < k$, given from the previous iteration:
\begin{align*}
    u_{k-1,j} = F(l_j \mid \overline{l}_{j-1}, do(\bar{a}_{k-1}), y_{k-1}=0), && q_{k-1,j} = F(\Tilde{y}_k \mid \overline{l}_{j-1}, do(\bar{a}_{k-1}), y_{k-1}=0).
\end{align*}
This can be achieved by noting that 
$$F(l_j \mid \overline{l}_{j-1}, do(\bar{a}_{k-1}),y_{k-1}=0) = F(l_j \mid \overline{l}_{j-1}, do(\bar{a}_{k}),y_{k-1}=0),$$ 
and 
$$F(\Tilde{y}_k \mid \overline{l}_{j-1}, do(\bar{a}_{k-1}),y_{k-1}=0) = F(\Tilde{y}_k \mid \overline{l}_{j-1}, do(\bar{a}_{k}),y_{k-1}=0),$$ 
which allows us to apply (\ref{eqn:ljyk}) to compute the required quantiles. 

We must then obtain 
$F(y_{k+1} \mid \overline{l}_{j}, do(\bar{a}_{k}), y_k=0)$ 
for each $j$ to use in the next iteration.  This can be done using (\ref{eqn:ylk}), 
and hence, we can move to the $(k+1)$th iteration. 

By the principle of induction, the statement is true for all $k\geq 0$.
\end{proof}



\section{Additional details of the illustration in Section \ref{sec:soul} }\label{sec:app_soul}
In our streamlined setup, we consider baseline variables $B =\{$Presence of CV disease$\}$ and time-varying covariate $L$ = $\{$HbA1c$\}$. We consider 7 ($K=6$) half-yearly visits leading to a $3.5$-year follow up periods. $A$ indicates the primary treatment, semaglutide and we assume perfect adherence so it is not time-varying. $S_k$ indicates the use of SGLT2-i at follow-up $k$. The time to first occurrence of MACE to be $T$ and $Y_k = I(T>k)$ being the binary survival status observed at each visit.  Specifically we choose that at baseline:
\begin{align*}
    B \sim \Bernoulli(0.7) && L_0 \sim \operatorname{Gamma}\!\left(k=(25+0.8B), \; \theta = 2.5\right)\\
    A \sim \Bernoulli(0.5) && S_0 \sim \Bernoulli(0.1).
\end{align*}
The choice of these parameter values are loosely based on the baseline characteristics of SOUL reported in \cite{mcguire2023effects}. 
For subsequent follow-up visits $k > 0$:
\begin{align*}
    S_k &\sim \Bernoulli(\expit(\alpha_0 + \alpha_1\,B + \alpha_2\,L_k +\alpha_3 \,A))\\
    L_k &\sim \operatorname{Gamma}(k=400(L_{k-1} + 0.2\,B - 0.5\,S_{k-1} - A_{k-1}), \; \theta =\frac{1}{400}). 
\end{align*}

We target a Cox marginal structural model given by:
\begin{align*}
    \lambda(\Tilde{y}_{k+1}\cmid do(A,\bar{S}_ k),Y_{k} = 0  ) = \lambda_0\exp\left(\beta_A \,A + \beta_S\,S_k + \beta_{AS}\,AS_k\right) =  g_k(A,S_k),
\end{align*}

This parameterization assumes a constant hazard rate between follow-up visits, influenced by the uses of semaglutide and SGLT2-i at the most recent visit. This implies that given survival at follow-up time $k$, the incremental survival time $\Tilde{Y}_{k+1}$ follows an exponential distribution with the rate equal to the hazard above:
\begin{align*}
    \Tilde{Y}_{k+1} \sim \operatorname{Exp}\big ( g_k(A,S_k) \big ). 
\end{align*}
If $\Tilde{Y}_{k+1} < 1$ then $Y_{k+1} =1$, and $0$ otherwise. 

To understand the impact of differential use of SGLT2-i on the estimation of treatment effect of the primary treatment, semaglutide, we create several scenarios with different prevalence of SGLT2-i in the placebo arm. We achieve this by varying the coefficients ($\alpha_0$,$\alpha_1$,$\alpha_2$ and $\alpha_3$) of the model for $S_k$. We also explored varying the efficacy of SGLT2-i. Table \ref{tab:sglt2_scenarios} below describes 4 levels of prevalence and 3 levels of efficacy of SGLT2-i. In total, we simulate 10 scenarios.

\begin{table}[ht]
\centering
\begin{adjustbox}{max width=\textwidth}
\begin{tabular}{ll}
\toprule
\midrule
\multicolumn{2}{l}{\textbf{SGLT2-i Prevalence Scenarios}} \\
\midrule
\textbf{Baseline} & $S_k \sim \text{Bernoulli}(\text{expit}(-9.5 + 0.5 B + 0.1 L_k - 1.5 A))$. \\
& Annual take-up: 10\% in placebo, 2.5\% in semaglutide group. \\
\textbf{High drop-in} & $S_k \sim \text{Bernoulli}(\text{expit}(-9 + 0.5 B + 0.1 L_k - 2 A))$. \\
& Annual take-up: 15\% in placebo, 2.5\% in semaglutide group. \\
\textbf{Low drop-in} & $S_k \sim \text{Bernoulli}(\text{expit}(-10 + 0.5 B + 0.1 L_k - A))$. \\
& Annual take-up: 5\% in placebo, 2.5\% in semaglutide group. \\
\textbf{No drop-in} &  $S_k = 0$ \\
\midrule
\multicolumn{2}{l}{\textbf{Effect Scenarios}} \\
\midrule
\textbf{Moderate effect} & $\beta_S = \log(0.9)$, $\beta_{AS} = 0.05$. \\
\textbf{Strong effect} & $\beta_S = \log(0.85)$, $\beta_{AS} = 0.1$. \\
\textbf{Weak effect} & $\beta_S = \log(0.95)$, $\beta_{AS} = 0.01$. \\

\bottomrule
\end{tabular}
\end{adjustbox}
\caption{Descriptions of SGLT2-i prevalence and efficacy scenarios. The prevalence scenarios vary the annual uptake of SGLT2-i in the placebo and semaglutide groups, while efficacy scenarios are defined by combinations of $\beta_S$ and $\beta_{AS}$, representing the isolated and interaction effects, respectively.}
\label{tab:sglt2_scenarios}
\end{table}

The values of $\beta_S$ are based on the findings of \cite{mcguire2021association}, who reported that SGLT2 inhibitors were associated with a reduced risk of MACE, with a hazard ratio of $0.90$ ($95\%$ confidence interval: 0.85--0.95) from a meta-analysis of six clinical trials. The interaction term represents the expectation that the combined effect of using both treatments is less than the total of their individual effects.

 For each simulated dataset, we fit the above Cox marginal structural model $\lambda(\Tilde{y}_{k+1}\cmid do(A,\bar{S}_ k),Y_{k} = 0  ) = \lambda_0\exp\left(\beta_A \,A + \beta_S\,S_k + \beta_{AS}\,AS_k\right) $ using stabilized inverse-probability-of-treatment weights \citep{hernan2000marginal}, calculated as
\begin{align}
    W_k = \prod_{j=0}^k \frac{\hat{p}(S_j \mid A, \bar{S}_{j-1}, Y_j = 0)}{\hat{p}(S_j \mid B, A, \bar{L}_j, \bar{S}_{j-1}, Y_j = 0)}.
\end{align}

\begin{remark}
        The denominator in the weight formula represents the correctly specified propensity score model under the assumed data-generating process, while the numerator stabilizes the weights. To avoid unintended confounding, the baseline covariates included in the numerator must also be adjusted for in the marginal structural model \citep{cole2008constructing}. In accordance with the model in (\ref{eqn:sim_MSM}), we only include $A$ in the numerator for stabilization.
\end{remark}

\section{\cite{seaman2023simulating} simulations}\label{sec:app_seaman}
We present an example that broadly replicates the setting described in \cite{seaman2023simulating}. We consider baseline variables $B$ and $C$, two time-varying covariates $L_k$ and $W_k$, and treatment $A_k$ at follow-up $k$. The time-to-event is $T$, with $Y_{k+1} = I(T<k + 1)$ indicating failure status observed at the end of the $k$th interval. The auxiliary continuous variable $\Tilde{ Y}_{k+1}$ is defined as previously. 
Specifically we choose that at baseline:
\begin{align*}
    B \sim \Exp(2) &&  L_0 \sim  \operatorname{Gamma}(k = 1 + 0.5 \,B + 0.5\,C, \; \theta =1)\\
   C \sim \Bernoulli(0.7)  && W_0 \sim \Bernoulli(\expit\{-0.2+0.5\,B + 0.5\,C\}).
\end{align*}
We specify that throughout the treatment assignment is influenced by $B$, $C$, $L_k$ and $W_k$ as: 
\begin{align*}
       A_k \sim \Bernoulli(\expit\{-1.5+0.5\,B + 0.5\,C + 0.5 \,L_k + 0.5\,W_k\}).
\end{align*}
For subsequent follow-up visits $k > 0$:
\begin{align*}
    L_k &\sim  \operatorname{Gamma}(k = 1 + 0.5 \,B + 0.5\,C + 0.1 \,L_{k-1} - 0.5 \,A_{k-1}, \; \theta =1)\\
    W_k &\sim \Bernoulli(\expit\{-0.2+0.5\,B + 0.5\,C + W_{k-1} - 0.6 \, A_{k-1}\}).
\end{align*}
We target a Cox marginal structural model given by:
\begin{align}
    \lambda(\Tilde{y}_{k+1}\cmid do(\bar{a}_k),c, y_k = 0) = \lambda_0\exp\left(\beta_C \,c + \beta_A\,a_{k} + \beta_{CA}\,c\,a_k\right), \label{eqn:sim_MSM}
\end{align}
where $\lambda_0 =  \exp(-2)$ is the baseline hazard. We set $\beta_C = -0.5$ and vary the values of $(\beta_A,\beta_{CA})$ to construct scenarios with different strengths of treatment: a strong effect scenario $(-0.5,0.3)$ and a weak effect scenario  $(0.2,0.1)$.
This parameterization assumes a constant hazard rate between follow-up visits, influenced by the treatment status at the most recent visit. This implies that given survival at follow-up time $k$, the incremental survival time $\Tilde{Y}_{k+1}$ follows an exponential distribution with the rate equal to the hazard above:
\begin{align*}
    \Tilde{Y}_{k+1} \sim \operatorname{Exp}(\lambda_0\exp\left(\beta_C \,c + \beta_A\,a_{k} + \beta_{CA}\,c\,a_k\right) ). 
\end{align*}
If $\Tilde{Y}_{k+1} < 1$ then $Y_{k+1} =1$, and $0$ otherwise. We set the total number of follow-ups $K$ to 6. 
We set the copulas describing the association between $L_j$ ($j = 0,\ldots,k$) and $Y_{k+1}$ to be Student t with parameter $\rho_L$ and 5 degrees of freedom, although each pairwise dependence could, in principle, be distinct. Similarly, we model the copulas capturing the dependence between $Y_{k+1}$ and each $W_j$ ($j = 0,\ldots,k$), given $L_k$, as Gaussian with correlation $\rho_W$, though again we note that different copulas could be specified for each time point. We vary the values of $\rho_L$ and $\rho_W$ to construct scenarios with different strengths of confounding. Table \ref{tab:demo2_scenarios} summarizes the combinations of treatment effect size and confounding levels considered. 

\begin{table}[ht]
\centering
\begin{adjustbox}{max width=\textwidth}
\begin{tabular}{ll}
\toprule
\midrule
\multicolumn{2}{l}{\textbf{Effect size}} \\
\midrule
\textbf{Large} & $(\beta_A,\beta_{CA}) = (-0.5,0.3) $ \\
\textbf{Small} & $(\beta_A,\beta_{CA}) = (0.2,0.1) $ \\

\midrule
\multicolumn{2}{l}{\textbf{Level of confounding}} \\
\midrule
\textbf{High} &  $(\rho_L,\rho_W) = (0.9 ,0.7) $ \\
\textbf{Medium} & $(\rho_L,\rho_W) = (-0.5,0.4) $\\
\textbf{Low} & $(\rho_L,\rho_W) = (-0.1,-0.2) $\\
\bottomrule
\end{tabular}
\end{adjustbox}
\caption{Parameter values of treatment effect size and confounding scenarios.}
\label{tab:demo2_scenarios}
\end{table}

We simulate data for sample size $n = \{1000, 2000, 5000\}$.

 We contrast two approaches for fitting the Cox marginal structural model specified in (\ref{eqn:sim_MSM}): stabilized inverse-probability-weighting and unweighted estimation. 
The first employs stabilized inverse probability of treatment weighting, with weights for subject $i$ at time step $k$ calculated as:
\begin{align*}
    \prod_{j=0}^k \frac{\hat{p}_j(A_j \mid C, \bar{A}_{j-1}, Y_j = 0)}{\hat{p}_j(A_j \mid B, C, \bar{L}_j,\bar{W}_j, \bar{A}_{j-1}, Y_j = 0)},
\end{align*}
where $\hat{p}_j(A_j \mid C, \bar{A}_{j-1}, Y_j = 0)$ and $\hat{p}_j(A_j \mid B, C, \bar{L}_j,\bar{W}_j, \bar{A}_{j-1}, Y_j = 0)$ are estimated using logistic regression models fitted to the survivors at time step $j$.

Table \ref{tab:demo2_bias} summarizes the bias of estimates from 500 replications, compared to corresponding Monte Carlo standard errors as a measure of  uncertainty. As the sample size increases, generally, the biases of estimates for $\beta_{C}$, $\beta_{A}$ and $\beta_{AC}$ obtained with inverse probability weights converge to zero, with their absolute values substantially smaller than the corresponding Monte Carlo standard errors. In contrast, the unweighted estimator exhibits considerable bias, particularly for the estimate of the treatment effect, $\beta_A$, which persists even at $n = 5000$. The magnitude of this bias far exceeds the Monte Carlo standard errors. The unweighted estimates are most biased under the high level of confounding by $L$ and $W$, since this is not adjusted for.


 \begin{table}[ht]
\centering
\begin{adjustbox}{width=\linewidth}
\begin{tabular}{rrrrrrrrrr}
\toprule
\multicolumn{1}{c}{\multirow{2}{*}{\makecell{\textbf{Level of} \\ \textbf{confounding}}}} & \multicolumn{1}{c}{\multirow{2}{*}{\textbf{Effect size}}} & \multicolumn{1}{c}{\multirow{2}{*}{$n$}} & \multicolumn{2}{c}{\textbf{$\beta_{C}$}} & \multicolumn{2}{c}{\textbf{$\beta_{A}$}} & \multicolumn{2}{c}{\textbf{$\beta_{AC}$}} \\
\cmidrule(lr){4-5} \cmidrule(lr){6-7} \cmidrule(lr){8-9}
& & & \multicolumn{1}{c}{IPTW} & \multicolumn{1}{c}{Unweighted} & \multicolumn{1}{c}{IPTW} & \multicolumn{1}{c}{Unweighted} & \multicolumn{1}{c}{IPTW} & \multicolumn{1}{c}{Unweighted} \\
\midrule
\textbf{High}&&$1000$ & $0.00 (2.18)$ & $-0.74 (1.46)$ & $-0.16 (2.26)$ & $2.75 (1.88)$ & $0.21 (2.99)$ & $0.22 (2.20)$ &  \\
&$\beta_A = -0.5$&$2000$ & $-0.07 (1.50)$ & $-0.74 (1.00)$ & $-0.13 (1.56)$ & $2.71 (1.25)$ & $0.24 (2.06)$ & $0.24 (1.51)$ &  \\
&&$5000$ & $-0.04 (1.13)$ & $-0.78 (0.67)$ & $-0.02 (1.03)$ & $2.89 (0.80)$ & $0.04 (1.46)$ & $0.11 (1.02)$ &  \\
\cmidrule{2-9}
&&$1000$ & $0.09 (2.26)$ & $-0.51 (1.60)$ & $0.27 (1.92)$ & $3.00 (1.56)$ & $-0.16 (2.76)$ & $0.07 (2.02)$ &  \\
&$\beta_A = 0.2$&$2000$ & $0.04 (1.64)$ & $-0.63 (1.15)$ & $0.07 (1.54)$ & $2.84 (1.15)$ & $-0.02 (2.04)$ & $0.21 (1.47)$ &  \\
&&$5000$ & $-0.04 (1.07)$ & $-0.73 (0.70)$ & $0.03 (0.91)$ & $2.81 (0.68)$ & $0.03 (1.26)$ & $0.31 (0.88)$ &  \\
\midrule
\textbf{Medium}&&$1000$ & $-0.13 (2.12)$ & $-0.77 (1.48)$ & $0.08 (2.40)$ & $2.66 (1.78)$ & $0.00 (3.12)$ & $0.20 (2.23)$ &  \\
&$\beta_A = -0.5$&$2000$ & $-0.01 (1.60)$ & $-0.79 (0.97)$ & $0.03 (1.72)$ & $2.53 (1.25)$ & $0.10 (2.29)$ & $0.39 (1.52)$ &  \\
&&$5000$ & $0.00 (1.04)$ & $-0.74 (0.60)$ & $-0.04 (1.09)$ & $2.58 (0.79)$ & $0.10 (1.52)$ & $0.30 (0.96)$ &  \\
\cmidrule{2-9}
&&$1000$ & $0.01 (2.34)$ & $-0.70 (1.56)$ & $0.28 (1.97)$ & $2.63 (1.54)$ & $-0.11 (2.90)$ & $0.30 (2.05)$ &  \\
&$\beta_A = 0.2$&$2000$ & $0.01 (1.73)$ & $-0.67 (1.04)$ & $0.06 (1.38)$ & $2.50 (1.08)$ & $0.01 (2.02)$ & $0.35 (1.37)$ &  \\
&&$5000$ & $-0.02 (1.12)$ & $-0.73 (0.68)$ & $0.05 (1.03)$ & $2.46 (0.66)$ & $0.05 (1.41)$ & $0.41 (0.86)$ &  \\

\midrule
\textbf{Low}&&$1000$ & $-0.09 (2.04)$ & $-0.11 (1.41)$ & $-0.05 (2.40)$ & $0.64 (1.75)$ & $0.12 (3.12)$ & $-0.12 (2.25)$ &  \\
&$\beta_A = -0.5$&$2000$ & $0.02 (1.57)$ & $-0.12 (0.97)$ & $0.02 (1.70)$ & $0.63 (1.22)$ & $0.09 (2.26)$ & $0.05 (1.50)$ &  \\
&&$5000$ & $-0.06 (1.07)$ & $-0.12 (0.65)$ & $-0.01 (1.11)$ & $0.70 (0.77)$ & $0.08 (1.47)$ & $-0.07 (0.94)$ &  \\

\cmidrule{2-9}

&&$1000$ & $0.14 (2.21)$ & $0.04 (1.44)$ & $0.27 (2.09)$ & $0.72 (1.54)$ & $-0.16 (2.85)$ & $-0.14 (1.92)$ &  \\
&$\beta_A = 0.2$&$2000$ & $-0.02 (1.75)$ & $-0.08 (1.05)$ & $0.01 (1.62)$ & $0.58 (1.11)$ & $0.12 (2.23)$ & $-0.01 (1.41)$ &  \\
&&$5000$ & $0.11 (1.03)$ & $-0.07 (0.60)$ & $0.10 (0.89)$ & $0.58 (0.64)$ & $-0.11 (1.32)$ & $-0.02 (0.81)$ &  \\

\bottomrule
\end{tabular}
\end{adjustbox}
\caption{Bias and Monte Carlo standard errors ($\times 10$) of estimates produced by inverse probability of treatment weighted (`IPTW') and unweighted Cox regressions, under different strengths of confounding, effect sizes, and sample sizes.}
\label{tab:demo2_bias}
\end{table}


Table \ref{tab:demo2_cov} presents coverage rates of the 95\% confidence intervals. Confidence intervals for the weighted estimator, computed using both sandwich variance estimation and bootstrap resampling, achieve coverage close to the nominal level, with slight undercoverage. In contrast, the sandwich-based confidence intervals for the unweighted estimator perform poorly, especially for the treatment effect estimate, $\beta_A$.

\begin{table}[ht]

\centering

\begin{adjustbox}{width=\linewidth}
\begin{tabular}{rrrrrrrrrrrr}
\toprule
\multicolumn{1}{c}{\multirow{2}{*}{\makecell{\textbf{Level of} \\ \textbf{confounding}}}} & \multicolumn{1}{c}{\multirow{2}{*}{\textbf{Effect Size}}} & \multicolumn{1}{c}{\multirow{2}{*}{$n$}} & \multicolumn{3}{c}{\textbf{$\beta_{C}$}} & \multicolumn{3}{c}{\textbf{$\beta_{A}$}} & \multicolumn{3}{c}{\textbf{$\beta_{AC}$}} \\
\cmidrule(lr){4-6} \cmidrule(lr){7-9} \cmidrule(lr){10-12}
& & & \multicolumn{1}{c}{W-S} & \multicolumn{1}{c}{W-B} & \multicolumn{1}{c}{UW} & \multicolumn{1}{c}{W-S} & \multicolumn{1}{c}{W-B} & \multicolumn{1}{c}{UW} & \multicolumn{1}{c}{W-S} & \multicolumn{1}{c}{W-B} & \multicolumn{1}{c}{UW} \\
\midrule
\textbf{High}& &1000 & 93.5\% & 94.3\% & 92.5\% & 93.8\% & 93.5\% & 60.3\% & 94.3\% & 93.8\% & 95.0 \% \\ 
&$\beta_A = -0.5$ &2000 & 94.8\% & 95.3\% & 89.8\% & 95.3\% & 93.3\% & 39.0\% & 95.3\% & 93.8\% & 95.5 \% \\ 
& &5000 & 92.3\% & 92.0\% & 75.5\% & 93.8\% & 94.5\% & 5.8\% & 94.0\% & 92.8\% & 93.5 \% \\ 
\cmidrule{2-12}
& &1000 & 93.5\% & 94.3\% & 93.0\% & 95.5\% & 95.5\% & 49.3\% & 94.5\% & 94.3\% & 93.0 \% \\ 
&$\beta_A = 0.2$ &2000 & 92.5\% & 93.0\% & 90.0\% & 93.8\% & 92.5\% & 26.5\% & 94.0\% & 93.5\% & 92.8 \% \\ 
& &5000 & 93.5\% & 94.0\% & 82.0\% & 94.8\% & 93.5\% & 1.0\% & 94.8\% & 94.3\% & 94.0 \% \\ 
\midrule
\textbf{Medium}& &1000 & 94.8\% & 94.8\% & 91.3\% & 94.3\% & 95.0\% & 64.0\% & 93.5\% & 93.3\% & 95.8 \% \\ 
&$\beta_A = -0.5$ &2000 & 94.3\% & 94.3\% & 88.8\% & 93.8\% & 93.8\% & 44.3\% & 95.0\% & 94.0\% & 95.8 \% \\ 
& &5000 & 97.5\% & 96.8\% & 83.0\% & 95.0\% & 94.8\% & 8.8\% & 94.3\% & 93.5\% & 92.8 \% \\ 
\cmidrule{2-12}
& &1000 & 93.3\% & 93.3\% & 91.8\% & 93.8\% & 93.5\% & 58.3\% & 92.0\% & 93.8\% & 94.8 \% \\ 
&$\beta_A = 0.2$ &2000 & 93.5\% & 94.8\% & 91.0\% & 96.5\% & 96.8\% & 36.3\% & 94.8\% & 95.8\% & 93.5 \% \\ 
& &5000 & 96.0\% & 94.5\% & 79.8\% & 94.5\% & 93.8\% & 6.5\% & 96.0\% & 95.0\% & 91.8 \% \\ 
\midrule
\textbf{Low}& &1000 & 93.5\% & 93.8\% & 94.0\% & 93.5\% & 93.3\% & 94.0\% & 93.5\% & 94.0\% & 94.5 \% \\ 
&$\beta_A = -0.5$ &2000 & 94.8\% & 95.3\% & 94.0\% & 95.8\% & 95.3\% & 92.5\% & 94.0\% & 94.5\% & 95.8 \% \\ 
& &5000 & 93.8\% & 94.3\% & 92.3\% & 95.3\% & 95.0\% & 86.0\% & 95.3\% & 93.8\% & 95.0 \% \\ 
\cmidrule{2-12}
& &1000 & 93.0\% & 94.8\% & 94.8\% & 93.3\% & 94.5\% & 92.8\% & 95.3\% & 95.8\% & 94.5 \% \\ 
&$\beta_A = 0.2$ &2000 & 92.0\% & 92.3\% & 94.0\% & 92.5\% & 93.8\% & 90.8\% & 92.3\% & 94.0\% & 94.0 \% \\ 
& &5000 & 96.5\% & 96.2\% & 96.0\% & 95.2\% & 95.2\% & 88.5\% & 95.0\% & 94.5\% & 96.5 \% \\

\bottomrule
\end{tabular}
\end{adjustbox}
\caption{Coverage rates for IPTW-Sandwich (W-S), IPTW-Bootstrap (W-B), and unweighted (UW) estimators of $\beta_{C}$, $\beta_A$, and $\beta_{AC}$ under different strengths of confounding, effect sizes, and sample sizes.}
\label{tab:demo2_cov}
\end{table}
In summary, the asymptotically unbiased nature of the inverse probability weighted estimates, compared to the biased unweighted estimates, further confirms that the simulated data were generated in accordance with the Cox model in (\ref{eqn:sim_MSM}), and that confounding by $L$ and $W$ was appropriately introduced.


\bibliographystyle{abbrvnat}
\bibliography{refs}
\end{document}